\documentclass{sig-alternate}
\usepackage{mathptmx} % This is Times font

\usepackage{fancyhdr}
\usepackage[normalem]{ulem}
\usepackage[hyphens]{url}
\usepackage[sort,nocompress]{cite}
\usepackage[final]{microtype}
\usepackage[keeplastbox]{flushend}
\usepackage[bookmarks=true,breaklinks=true,letterpaper=true,colorlinks,citecolor=blue,linkcolor=blue,urlcolor=blue]{hyperref}

\usepackage{mathptmx} % This is Times font
\usepackage{array}
\usepackage{pifont}
\newcolumntype{P}[1]{>{\centering\arraybackslash}p{#1}}
\usepackage{xcolor}
\usepackage{amsmath,amssymb,amsfonts}
\usepackage{algorithmic}
\usepackage{graphicx}
\usepackage{textcomp}
\usepackage{fancyhdr}
\usepackage[normalem]{ulem}
\usepackage[hyphens]{url}
\usepackage{svg}
\usepackage{subcaption}
\usepackage{dblfloatfix}    % To enable figures at the bottom of page
\usepackage{soul}
\usepackage{xspace}
\usepackage{braket}
\usepackage{multirow}
\usepackage[inline]{enumitem}
\usepackage{caption}
\usepackage{authblk}
\usepackage{booktabs}
\usepackage{pifont}

\newcommand{\ignore}[1]{}

\newcommand{\tin}{\text{in}}
\newcommand{\tout}{\text{out}}

\newcommand{\epsDC}{\varepsilon_{\text{DC}}}
\newcommand{\epsPhi}{\varepsilon_{\phi}}

\usepackage{tikz}

\newcommand{\cmark}{\textcolor{green}{\ding{52}}}
\newcommand{\xmark}{\textcolor{red}{\ding{55}}}

\newcommand{\ElectroPhotoAcc}{ADEPT\xspace}
\newcommand{\PhotonicCore}{photo-core\xspace}
\newcommand{\PhotonicCores}{photo-cores\xspace}

\newcommand{\ElectronicAsic}{digital electronic ASIC\xspace}

\newcommand{\SystolicArray}{SA\xspace}
\newcommand{\SystolicArrays}{SAs\xspace}

\fancypagestyle{firstpage}{
  \fancyhf{}
  
  \fancyfoot[C]{\thepage}
}

\title{\Large{\textbf{An Electro-Photonic System for Accelerating Deep Neural Networks}}\vspace{-0.4in}}

\author[1]{Cansu Demirkiran\thanks{cansu@bu.edu}}
\author[1]{Furkan Eris}
\author[2]{Gongyu Wang}
\author[2]{Jonathan Elmhurst}
\author[2]{Nicholas Moore}
\author[2]{Nicholas C. Harris}
\author[2]{Ayon Basumallik}
\author[3]{Vijay J. Reddi}
\author[1]{Ajay Joshi}
\author[2]{Darius Bunandar}
\affil[1]{Boston University, $^2$ Lightmatter, $^3$ Harvard University}
\begin{document}
\maketitle
\thispagestyle{firstpage}
\pagestyle{plain}

%%%%%% -- PAPER CONTENT STARTS-- %%%%%%%%

\begin{abstract}
The number of parameters in deep neural networks (DNNs) is scaling at about 5$\times$ the rate of Moore's Law. 
To sustain this growth, photonic computing is a promising avenue, as it enables higher throughput in dominant general matrix-matrix multiplication (GEMM) operations in DNNs than their electrical counterpart.
However, purely photonic systems face several challenges including lack of photonic memory and accumulation of noise. In this paper, we present an electro-photonic accelerator, ADEPT, which leverages a photonic computing unit for performing GEMM operations, a vectorized digital electronic ASIC for performing non-GEMM operations, and SRAM arrays for storing DNN parameters and activations. 
In contrast to prior works in photonic DNN accelerators, we adopt a system-level perspective and show that the gains while large are tempered relative to prior expectations. 
Our goal is to encourage architects to explore photonic technology in a more pragmatic way considering the system as a whole to understand its general applicability in accelerating today's DNNs. 
Our evaluation shows that \ElectroPhotoAcc can provide, on average, 5.73$\times$ higher throughput per Watt compared to the traditional systolic arrays (SAs) in a full-system, and at least 6.8$\times$ and $2.5\times$ better throughput per Watt, compared to state-of-the-art electronic and photonic accelerators, respectively. 
\end{abstract}
\section{Introduction}
\label{sec:intro}

Deep neural networks (DNNs) have shown to perform impressive human-like tasks in a range of applications including image and video processing~\cite{resnet}, diagnostic medical imaging~\cite{unet}, speech recognition~\cite{rnnt}, and conversational AI~\cite{devlin2019bert}.
OpenAI's study shows that modern DNN computational requirements have increased 300,000$\times$ from AlexNet (2012) to AlphaGo Zero (2018). 
This general trend is projected to continue as newer and larger DNN models emerge ever so often~\cite{amodei_2020}.

Consequently, a variety of solutions have been developed to support the growing compute requirements. 
These solutions include massively-threaded graphics processing units (GPUs)~\cite{9361255, 6865444, deepmon}, field-programmable gate arrays (FPGAs)~\cite{lecun_fpga_2011, fpga-large-scale, massive-fpga2009}, and specialized application-specific integrated circuits (ASICs)~\cite{DianNao, DaDianNao,neuflow}. 
While these solutions provide significant architectural and performance benefits for DNN execution, they are based on CMOS transistors---devices that no longer scale in area or energy consumption according to Moore's Law and Dennard Scaling~\cite{mooreslaw}.

As an alternative, there is growing interest in using photonic computing architectures for meeting the computational demands of DNNs.
The idea of computing with light is not new and has been explored since the 1960s~\cite{Reimann1965,Bell:86,Caulfield:87}.
The advent of integrated photonics, in particular silicon photonics, which has seen widespread integration in commercial CMOS foundries alongside transistors on 300-mm wafers~\cite{Giewont:2019} has further propelled research in photonic computing.
However, limitations around photonic information storage (no photonic memory) and weak photon-photon nonlinearities (no photonic transistor) make it difficult---if not impossible---to design a general-purpose fully-photonic computing architecture.
Prior art leverages the highly parallel and efficient linear transformations enabled by photonics to build specialized DNN accelerators with orders of magnitude improvements in speed and energy efficiency when computing GEMM and convolution operations~\cite{CoherentNanophotonic2017,albireo,deapcnn,pixel,dnnara,Wetzstein:2020wd,Shastri:2021td}, which accounts for more than 90\% of the total number of operations within a DNN network~\cite{CHEN-dnn-acc-survey}.

In this paper, we seek to calibrate the expectations of the photonic GEMM technology with respect to building a complete system, including the photonic and non-photonic components needed to make it all work. We set out to answer two {key questions}. First, \textbf{given that photonic accelerators still need electronics (for control, data storage, and nonlinearities), how do we build a complete electro-photonic accelerator architecture that is not bottlenecked by the slower electronics?} To answer this question, we present the microarchitecture of an electro-photonic accelerator called ADEPT, where we match the throughput of the electronic and photonic components.
ADEPT comprises of high-throughput photo-core(s), various data converters, custom vectorized electronic digital ASIC, and large electronic SRAM arrays.
The photo-core is a scalable and highly-efficient photonic tensor core containing Mach-Zehnder Interferometers (MZIs) for GEMM operations.
In the photo-core, we adopt a weight stationary (WS) approach where the weight matrix is programmed into the MZI array.
The inputs are routed, one vector at a time, through digital-to-analog (D-A) converter, electrical-to-optical (O-E) converter, the MZI array, optical-to-electrical (O-E) converter and analog-to-digital (A-D) converter.
While the photo-core can handle GEMM operations (over $90\%$ of the overall DNN operations), DNNs rely on a non-trivial amount of non-GEMM operations that are executed in the electrical domain. 
To match the throughput of the photonic and electronic components, we architect a highly-vectorized electronic digital ASIC with multiple digital lanes, where each lane supports basic arithmetic operations that can be used for building more complex non-GEMM operations. 
To efficiently orchestrate the operations and maximize the performance of ADEPT, we pipeline GEMM and non-GEMM operations and use an efficient buffering scheme to minimize DRAM access overhead. Finally, we evaluate ADEPT in the context of a full system to understand the big picture.

The second question we set out to answer is \textbf{(2) how much are the electro-photonic accelerator systems better than purely electronic accelerator systems, when we consider the system as a whole, i.e., accelerator + memory + host processor + communication, running practical real-world applications?} To answer this question, we perform a head-to-head comparison of \ElectroPhotoAcc with electronic systolic arrays (\SystolicArrays) in terms of the full system throughput (in inferences per second or IPS), power efficiency (in IPS/W), and power-area efficiency (in IPS/W$\cdot \text{mm}^2$).
We use the following three state-of-the-art neural networks from the MLPerf datacenter inference benchmarks~\cite{mlperf} that represent a wide range of operations: ResNet-50~\cite{resnet} for image classification on the ImageNet dataset ~\cite{imagenet}, BERT-large~\cite{devlin2019bert} for natural language processing (NLP) on the SQuAD v1.1 ~\cite{squad} question-answering dataset, and RNN-T~\cite{rnnt} as an LSTM-based speech recognition network on the LibriSpeech~\cite{librispeech} speech audio dataset. 
Our analysis shows that, compared to \SystolicArrays, \ElectroPhotoAcc provides \textbf{4.89}$\times$, \textbf{3.24}$\times$ and \textbf{9.06}$\times$ better power efficiency
for the full system for ResNet-50, BERT-large and RNN-T networks, respectively.
Compared to the state-of-the-art electronic accelerators, \ElectroPhotoAcc performs at least \textbf{6.8}$\times$ better in terms of IPS/W.
In addition, we perform a detailed comparison between ADEPT and current state-of-the-art photonic accelerators. 
Our analysis shows that compared to state-of-the-art photonic accelerators~\cite{albireo, dnnara, holy-light}, ADEPT can provide more than \textbf{2.5}$\times$ better power efficiency for the same batch size and more than \textbf{8.3}$\times$ better power efficiency when the maximum batch size is used.

In summary, our work is the first to emphasize the importance of considering the entire system to understand the real benefit of the photonic GEMM cores for DNN inference. 
Our study shows that the impact of the electronic components in an electro-photonic accelerator system is not negligible. 
However, while an electro-photonic system may be bound by Amdahl's Law, it is still feasible---and beneficial---to build a balanced electronic-photonic system that leverages the highly-efficient photonic computing medium.
Our work aims to provide practical insights to the community and to encourage architects to explore photonic technology in a more pragmatic way without ``missing the forest for the trees''. 
Broadly, we show that while using photonics technology for computing is promising, claims of tera-inferences per second are not realistic when considering the system as a whole.
\section{Background \& Related Work}
\label{sec:bk-rel-wk}

This section provides an overview of how photonic devices perform GEMM operations and why performing nonlinearities in photonic systems is challenging. We then detail state-of-the-art photonic accelerators to demonstrate how our work sheds new insights, specifically from the perspective of a complete system rather than just the accelerator.

\subsection{GEMM Operations using Photonic Devices}
\label{sec:mvm}
Photonic computers can efficiently perform certain operations such as matrix-vector multiplication (MVM). 
A photonic core that performs MVM operations can be architected using an array of MZIs.
An MZI is a configurable photonic device that controls the interference of two light beams by adjusting the relative phase shift between the beams.
A simple MZI consists of two directional couplers and a differential phase-shift in between.
The transfer function of an MZI is represented by a $2\times2$ orthogonal matrix:
\begin{equation}
U(2) = \begin{bmatrix}
\sin \phi & \cos \phi\\
\cos \phi&- \sin \phi\\
\end{bmatrix},
\label{eq:mzi}
\end{equation}
where $\phi$ is the phase difference between the two internal arms of the MZI.
Similarly, MZIs can be used as an attenuator for scaling a single value when one arm is blocked.
In silicon photonics, the phase difference $\phi$ is achieved by delaying light in one arm using various mechanisms, including the thermo-optic effect ($\sim100$s kHz bandwidth)~\cite{Harris:14}, mechanical effect ($\geq$ MHz bandwidth)~\cite{noems}, and electric-field induced electro-optic effect ($\geq$ GHz bandwidth)~\cite{Timurdogan:2017wh}.

To perform an MVM using an MZI array, we first need to program the matrix into the MZIs as phase values.
Fig.~\ref{fig:sys-arch}-a shows an example of programming a $3\times3$ matrix $M$ into a $3 \times 3$ MZI array in Fig.~\ref{fig:sys-arch}-b. 
The matrix $M$ is first decomposed into the three matrices through the singular value decomposition (SVD), i.e., $M = U\Sigma V^T$, where $U$ and $V^T$ are $m \times m$ orthogonal matrices and $\Sigma$ is a diagonal matrix of singular values. 
The larger $U(m)$ and $V(m)^T$ orthogonal matrices (with $m > 2$) are composed by tiling $m(m-1)/2$ MZIs in a rectangular pattern~\cite{Clements:16,Harris:18}. 
Next, the phases $(\phi_U, \phi_{\Sigma}, \phi_{V^T})$ needed to program in the matrices $U$, $\Sigma$, and $V^T$ are computed by using the phase decomposition algorithm~\cite{Clements:16}.
The phase decomposition algorithm is an algorithm similar to QR decomposition that breaks a large orthogonal matrix $U$ into a series of $2\times2$ orthogonal matrices acting on different input rows.
Finally, a total of $m^2$ phase values---equal to the number of elements in $M$---are programmed into the array to create the matrix $M$.

An MVM between a matrix $M$ and a vector $v_{\tin}$ can then be achieved by (1) programming the matrix $M$ in the array of MZIs; (2) encoding the vector $v_{\tin}$ in the amplitude and phase (0 or $\pi$ for sign) of the optical signals entering the array; and (3) obtaining the resulting vector $v_{\tout} = M\cdot v_{\tin}$ at the output of the array. When the vector $v_{\tin}$ is inserted at GHz rate, a $100 \times 100$ array enables us to perform linear operations at $10$ Tera Operations per Second (TOPS). 
A GEMM operation consists of a series of MVM operations. GEMM between two matrices can be achieved by encoding one matrix in the MZI array and by sending the other matrix through the array as optical signals---one vector at a time.
\begin{table*}[ht]
    \centering
    \caption{Comparison against other photonic accelerators.}
    \label{table:photonic-acc}
    \begin{tabular}{cccccccccccc}
        \toprule
        {} & {} &  \multicolumn{3}{c}{Non-Photonic Components and Metrics Considered} & \multicolumn{3}{c}{Benchmarks}\\
        \cmidrule(lr){3-5}
        \cmidrule(lr){6-8}
        Accelerator & Optical Element & Non-GEMM & On-chip Memory & Off-chip Memory& CNN & NLP & RNN\\
        \midrule
        \textbf{ADEPT} & MZI  & \cmark & \cmark & \cmark & \cmark & \cmark & \cmark \\
        Albireo\cite{albireo} & MRR+MZI& \xmark & \cmark & \xmark & \cmark & \xmark & \xmark \\
        PIXEL\cite{pixel} & MRR+MZI & \cmark & \xmark & \xmark & \cmark & \xmark & \xmark \\
        PCNNA\cite{pcnna} & MRR & \xmark & \xmark & \xmark & \cmark & \xmark & \xmark \\
        DNNARA\cite{dnnara} & MRR  & \cmark & \cmark & \xmark & \cmark & \xmark & \xmark \\
        Holy-Light\cite{holy-light} & MRR & \cmark & \cmark & \xmark & \cmark & \xmark & \xmark \\
        \bottomrule
    \end{tabular}
\end{table*}
\subsection{Nonlinear Operations}
\label{sec:backgr-nonlinear}
Any nonlinear operation (e.g., nonlinear activation functions or conditional if-else statements) on the optical electromagnetic (EM) waves requires the use of nonlinear optical media~\cite{Jackson:100964}. 
Nonlinear optical activation function has previously been demonstrated using laser-cooled atoms, which absorb light up to some saturation intensity (higher intensity light is absorbed more)~\cite{Zuo:19}.
Saturable absorbers, where the amount of light absorbed decreases with increasing light intensity, have also been proposed as optical nonlinear activations~\cite{CoherentNanophotonic2017,Bao:2011uv}. 
However, the practical implementation of these nonlinear optical activations remains challenging, especially since (1) they have not been miniaturized, and (2) repeated usage of the nonlinear activation function will decay the signal quickly.

Amplification-based nonlinear functions made out of semiconductor optical amplifiers (SOAs) in III-V materials, e.g., InP and InGaAs, can combat the loss described above~\cite{Roelkens:2007}.
In principle, an optical DNN accelerator can be built in the III-V platform itself~\cite{Shi:2020}, but researchers still prefer to use silicon photonics as it has been monolithically integrated with the CMOS transistors~\cite{Giewont:2019} needed for controlling the photonic components. 
Packaging the III-V module with a silicon photonics module poses a challenge to its feasibility. 
Even when a practical packaging solution is available, the amount of power needed to maintain the optical signal throughout the entire inference will increase exponentially with the number of neural network layers.
We therefore conclude that optical nonlinearities are impractical today, and we choose to architect a system that performs these nonlinearities electronically.

\subsection{State-of-the-Art Photonic Accelerators}

Previous works have proposed several photonic tensor core architectures isolated from the system surrounding the cores~\cite{Feldmann:2021tm, Xu:2021vb, CoherentNanophotonic2017, Wetzstein:2020wd,Shastri:2021td}.
While the performance numbers are impressive, these accelerators need to be viewed through the lens of a practical system.
Table~\ref{table:photonic-acc} presents the state-of-the-art photonic accelerator architectures and sets the stage to discuss how one needs to systematically take a full-system view. 

\subsubsection{Optical Elements}  
Accelerators in Table~\ref{table:photonic-acc} are based primarily on microring resonators (MRRs), which are typically smaller (with a dimension of $\sim 10$ $\mu$m) than MZIs (with a dimension of $\sim 100$ $\mu$m) and can provide a better power and area efficiency~\cite{mzi-vs-mrr}.
MRRs require stabilization circuits for operations that have been demonstrated for communication~\cite{PadmarajuBergman2014, 8310328}.
However, for computation, the bit precision of this circuitry would need to be higher to support the precision of the computation (more than the 1 or 2 bits required for NRZ or PAM-4 keying, respectively).
As such, the stabilization circuitry will consume more area and power than what has been previously demonstrated.
A single MZI has been shown to achieve an extremely high extinction ratio (ER, which is a measure of how precise the light signals can be modulated by the photonic device) of greater than 60~dB~\cite{mzi-high-extinction}.
In contrast, the ER of MRRs is determined by how closely critical coupling can be achieved which can be limited by the MRR's thermal stability~\cite{rings-bogaerts}. 
State-of-the-art demonstrations of a single MRR have their measured ER at $< 25$~dB~\cite{microring-high-extinction}. 
Hence, using MZIs, as we do in ADEPT, is a more scalable and practical solution, and Shen et al.~\cite{CoherentNanophotonic2017} have demonstrated their applicability for DNN acceleration.

\subsubsection{Compute vs. Memory} 
Small on-chip caches (on the order of hundreds of KBs used by previous works~\cite{albireo, pixel, dnnara}) cannot hold large DNN models, input/output data and intermediate data at the same time, and so will need frequent off-chip memory accesses---which will stall the photonic core.
Similarly, non-GEMM operations should be performed fast enough not to throttle down the high-throughput photonic core.
Therefore, all the electronic components in the accelerator should be architected carefully and analyzed in detail to make fair conclusions about the photonic technology.
Unfortunately, the studies of the non-photonic components in the previous works have been limited.
In our work, we provide a complete system-level analysis in terms of power and latency including the non-photonic arithmetic units for non-linear operations, data conversion circuits, die-to-die interconnect, and on-chip and off-chip memory.

\subsubsection{Benchmarks} 
Prior photonic accelerators are either specifically designed for CNNs or report results only for CNNs.
While these accelerators perform well for convolution operations, most are under-utilized and perform poorly for linear layers.
Moreover, several of them use old and small neural networks that do not stress the memory system as much as the state-of-the-art neural networks.
Additionally, non-CNN networks are typically richer in terms of the variety of operations---which makes the system perspective even more important.
Given that non-CNN networks are being more commonly used in the recent years, focusing on only CNNs provides a limited perspective on using photonic cores for DNN acceleration. 
\ElectroPhotoAcc is the first photonic accelerator work to report results for non-CNN networks, particularly with BERT-Large and RNN-T that contain a wider range of operations than CNNs.

Broadly, while previous works are helpful towards understanding the raw capability of photonic compute cores, our key take-away message here is that it is not just about the raw compute capacity of photonic cores; instead, it is important to look at the system as a whole and understand the general applicability and true benefits of photonic technology in AI.

\section{Full-system Architecture}
\label{sec:full-sys-arch}
\begin{figure*}[t]
    \centering
    \includegraphics[width=\linewidth]{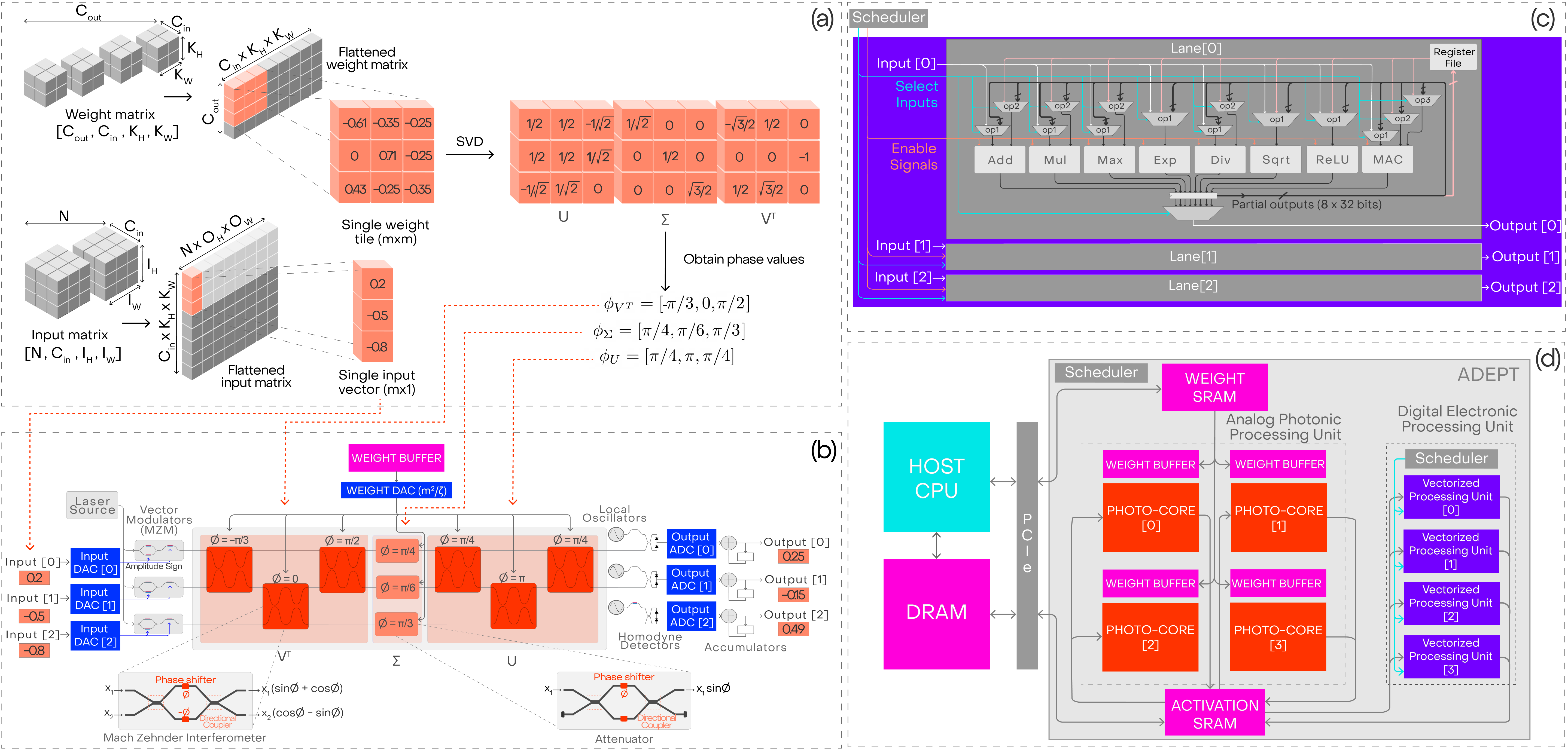}
    \caption{\textbf{Diagram showing different components of ADEPT and how operations are performed.} (a) Example GEMM operation in the photo-core.
    (b) Programming input and weight matrices into the photo-core.
    The $m\times m$ (here $m=3$ as an example) photo-core consists of $2 \times m (m-1)/2 = 6$ MZIs (for $U$ and $V^T$) and 3 attenuators (for $\Sigma$).
    (c) Microarchitecture for a single digital electronic vectorized processing unit. 
    The unit comprises $m=3$ digital lanes, each consisting of arithmetic units to perform non-GEMM operations. 
    (d) Full system architecture including the host CPU, the DRAM, and ADEPT---interconnected using a PCI-e interface. As an example, we show four photo-cores and four vectorized processing units.}
    \label{fig:sys-arch}
\end{figure*}
Our work focuses on understanding the implications of a complete electro-photonic system consisting of a host CPU, DRAM, PCI-e bus and the electro-photonic accelerator ADEPT (see Fig.~\ref{fig:sys-arch}-d).
\ElectroPhotoAcc is connected to the host CPU and DRAM through a PCI-e bus.
Host CPU handles the compilation and any other operations required by the DNN model that can be performed offline including pre/post processing (e.g., resizing, decoding, etc.) and precomputation of the phase values for the MZIs. 
The inference is then performed fully in ADEPT without any interference from the host CPU. 
In this section, we introduce the ADEPT (micro)architecture, present optimizations that allow it to be efficiently integrated into a full system, and describe the compilation flow, so that we can do a full evaluation of an electro-photonic system. 

\subsection{\ElectroPhotoAcc Architecture}

ADEPT is an electro-photonic accelerator that contains an analog photonic computing unit for GEMM operations, a custom digital electronic vectorized processing unit for non-GEMM operations, and memory units for storing weight and activation data.

\subsubsection {Analog Photonic Computing Unit}
\label{sec:photonic-unit}
The photonic computing unit in \ElectroPhotoAcc is an analog unit designed to perform MVM operations that can eventually be composed into a GEMM operation (see Section~\ref{sec:mvm}). 
The unit consists of a set of vector modulators (Mach-Zehnder Modulators (MZMs)), an array of MZIs, photo-detectors, analog-to-digital converters (ADCs), and digital-to-analog converters (DACs) (see Fig.~\ref{fig:sys-arch}(b)).
We refer to the unit without ADCs and DACs as the \PhotonicCore.

GEMM operations in DNNs (e.g., in the fully-connected (FC) layer and the 2D convolution layer) typically involve a multiplication between a weight tensor and an input tensor.
The input and weight matrix shapes vary for each layer in a DNN, but the photo-core has a fixed size of $m\times m$. 
Therefore, matrices bigger than $m \times m$ are divided into $m \times m$ sized submatrix tiles and loaded into the photo-core one-by-one.

We adopt a WS dataflow in the photo-core, where the weight matrix is programmed into the MZI array and the input vector is encoded in the optical signals.
Fig.~\ref{fig:sys-arch}(a) shows a simple example of this step.
First, the input and weight matrices are flattened if necessary (for 2D convolutions) using `im2col' pre-processing~\cite{gemm-lowmem}, and the weight matrix is broken into submatrix tiles.
Each weight tile is then decomposed into two orthogonal matrices ($U$ and $V^T$) and a diagonal matrix of singular values ($\Sigma$) using SVD.
Next, each of the three matrices is decomposed into their respective phase values ($\phi_U, \phi_\Sigma, \phi_{V^T}$) using the phase decomposition algorithm~\cite{Clements:16}.
Tiling and decomposition are performed \emph{only once upfront} for each weight submatrix tile in the host CPU.
Therefore, it does not introduce a latency overhead during the inference.
Importantly, the total number of phase values is equal to the number of elements in the weight tile. 
Therefore, the memory footprint required for storing the decomposed parameters is the same as that for storing the original tile.
The phase values obtained from above can be directly programmed into the MZI array, as shown in Fig.~\ref{fig:sys-arch}(b). 

Fig.~\ref{fig:sys-arch}(b) also shows an example of how the input and output vectors are programmed and read out, respectively.
Each element of the input vector is programmed using an MZM and a phase shifter which encode the amplitude and the sign (0 or $\pi$) of the input optical signals, respectively.
The output vector ($m \times 1$) of the MVM operation (both amplitude and sign) is detected using $m$ coherent detectors with the help of a local oscillator.
The resulting photocurrent is eventually converted into digital bits, using 8-bit ADCs. 
Partial results are accumulated digitally to construct the final outputs, which are then stored back into the activation SRAM.
The input and the weight DACs are chosen to be 10-bit and 12-bit precise, respectively, which are adequate to guarantee 8-bit precise outputs (see Section~\ref{sec:precision}).

In the WS approach, the weight values of a tile are first transferred from the weight SRAM into the weight buffer.
Data from the weight buffer can be programmed into the \PhotonicCore at a rate limited by the modulation mechanism of the MZIs---during which, the \PhotonicCore is inoperable.
This overhead is unavoidable but it is fairly small $\sim10$~ns~\cite{mi10010051}.
Once the tile is loaded into the photo-core, the values are maintained in the MZI array while all input vectors that need to be multiplied with this particular tile are fed into the photo-core vector-by-vector.
WS enables us to amortize the power and latency costs of programming the MZI array.
One can also consider an input stationary (IS) approach where the input matrix, instead of the weight matrix, is programmed into the MZI array.
However, the current architecture of the \PhotonicCore prohibits the IS approach because the input matrix of a DNN layer is the output of the previous layer and is computed at runtime. Therefore, SVD and phase decomposition algorithms, both of which have the same computational complexity of a GEMM operation, will also need to be performed on the input matrix to be programmed into the array at runtime instead of being performed offline.
Similarly, output stationary (OS) is not feasible in this architecture because it would require updating the values programmed into the MZIs each cycle.

\subsubsection {Digital Electronic Processing Unit}
\label{sec:electronic-unit}
Although more than 90\% of the operations are GEMM operations, a non-trivial amount of non-GEMM operations must also be performed as part of DNN inference.
These operations include element-wise non-linear operations (e.g., ReLU, GELU, and sigmoid); reduction operations (e.g., softmax and max-pool); batch and layer normalizations; and element-wise multiplication and addition (e.g, bias).
As discussed in Section~\ref{sec:backgr-nonlinear}, these non-GEMM operations are more effectively performed in the digital domain instead of the analog domain.

To maintain the balance between the analog and digital parts of ADEPT, within the \ElectronicAsic, we use the same number of vectorized processing units as the photo-cores. 
The microarchitecture of a single vectorized processing unit is shown in Fig.~\ref{fig:sys-arch}(c).
In each vectorized processing unit we use the same number of lanes as the number of optical lanes (channels) in one \PhotonicCore such that output of each optical lane in the \PhotonicCore is fed to one lane in the vectorized processing unit via the activation SRAM.
Each lane has separate units for multiplication, addition, division, max, square root, and exponential operations (each 32-bit) that enable the system to complete the wide variety of non-GEMM operations.
These arithmetic units are implemented as custom digital CMOS circuits.
All lanes in the vectorized processing unit can operate in parallel and can be pipelined for non-GEMM operations that require multiple arithmetic operations.
Each arithmetic unit uses a multiplexer to choose the input from 1) the activation SRAM, 2) the output of the arithmetic units, or 3) the register files of the vectorized unit, as operands.
Here the register files (64 KB each) are used to store the constants (which are loaded up front) for the non-GEMM operations or the outputs of the arithmetic units. 
Multiplexers are controlled by a scheduler that decides when each arithmetic operation is used.
The outputs of \ElectronicAsic are written back to the activation SRAM---to be used in the next layer of the DNN.

To extract the maximum performance from \ElectroPhotoAcc, we need to match the throughput of the \PhotonicCore and the \ElectronicAsic.
It is, however, challenging to design a digital ASIC that can operate above 2 GHz. 
Hence, we use $n$ logical units in parallel for each operation within the individual vector lane.
Each unit operates at $1/n$ times the clock frequency $f_c$ of the \PhotonicCore (each offset by $1/f_c$ to one another) to match the throughput of \PhotonicCore. 

\subsubsection {Data Movement and Storage}
\label{sec:memory}
\ElectroPhotoAcc utilizes two separate SRAM units: one for input/output activations and one for weights.
The SRAM units can transfer data between each other through direct memory access (DMA) and communicate with the host and DRAM through the PCI-e fabric. 
The two SRAM units are separated because, generally, a dichotomy exists between the activations and the weights, and data transfer between them is not frequent.
The activation SRAM is used to store both input and output activations because effectively, the output of one layer is the input of the next layer.
At runtime, both the photo-core and the digital electronic ASIC read and write a vector of size $m$ (the size of the photo-core) from and to the activation SRAM. 
We use separate dedicated read/write ports in the activation SRAM for the photo-core and the digital electronic ASIC. 

Transferring a complete weight tile ($m\times m$) from weight SRAM to \PhotonicCore in one step requires a large SRAM bandwidth. 
In contrast, transferring one vector at a time requires a large latency in between tiles. 
Hence, we use a weight buffer for each photo-core as an intermediate stage. 
We load the tile for the next set of GEMM operations into the weight buffer, while the \PhotonicCore is performing GEMM operations with the current weight values.
The data from the weight buffer is then programmed into the \PhotonicCore in $\sim10$~ns~\cite{mi10010051}, minimizing the latency in between consecutive tiles in the photo-core and increases the photo-core's overall utilization and the system throughput.

\subsubsection{Numerical Precision}
\label{sec:precision}
Maintaining the numerical precision of an entire DNN computation is one of the main challenges of computing with an analog photonic core: the numerical precision of the output vector $v_{\tout}$ is limited by how well one can encode the input vector $v_{\tin}$ and the matrix $M$.
The errors in the three quantities are related as follows: $\Delta v_{\tout}^2 = \Delta v_{\tin}^2 + \Delta M^2$.
The error of the input vector encoding $\Delta v_{\tin} \leq 2^{-b_{\tin}}$ is quantified by the bit precision of the input DACs, $b_{\tin}$.
Similarly, the output vector is captured by ADCs with bit precision of $b_{\tout}$ where $\Delta v_{\tout}$ must be $\leq 2^{-b_{\tout}}$.

Phase encoding error ($\epsPhi$) and directional coupler splitting error ($\epsDC$) mainly contribute to the error of the matrix $\Delta M$.
The depth of the photonic circuits in $M$ grow as $O(m)$, and splitting errors cascade as light propagates down the mesh. A na\"{i}ve programming of the phases gives $\Delta M^2 \sim O(m \epsPhi^2 + m \epsDC^2)$~\cite{Clements:16}.
However, more sophisticated error-corrected programming strategies~\cite{hwcorrection:2021,hamerly2021accurate,hamerly2021stability} can achieve a better scaling with respect to the errors, such that $\Delta M^2 \sim O(m \epsPhi^2 + m^2 \epsDC^4)$, which is advantageous when $\epsDC \leq m^{-1/2}$. Taking $\epsDC < 0.1~\%$ (measured in our fabricated wafers), the precision of the output vector can be maintained up to $\sim$8 bits for matrices up to size $256\times256$ if the input and the weight DAC bit precisions are 10 and 12 bits, respectively. 
\subsection{Optimizations}
In this section, we explain the optimizations that help us efficiently orchestrate the operations in ADEPT, reduce the latency overhead caused by the non-GEMM operations and data transfers, and maximize the system performance. 

\label{sec:optimizations}
 
\subsubsection{Pipelining Operations}
\label{sec:pipelining}
We pipeline GEMM and non-GEMM operations in \ElectroPhotoAcc. Specifically, once \emph{an output vector} (after accumulating the partial output results) of a GEMM operation has been generated, that output vector is immediately sent to the digital electronic ASIC for non-linear operations.
Therefore, non-GEMM operations begin without the need to wait for the whole GEMM operation to be completed.

In addition, more than one layers including non-GEMM operations can follow one another, or one layer may need to use more than a single logical unit.
We further optimize ADEPT by pipelining these non-GEMM operations in the digital electronic ASIC.
For example, the softmax layer uses the exponential unit, the max unit, and the multiplication unit. 
While one element is using the exponential unit, the previous output of the exponential unit uses the max unit. 
As a result, as long as the data dependency is preserved, different non-GEMM operations or different steps using different arithmetic units in the digital electronic ASIC within a non-GEMM operation can be parallelized and pipelined. 

\subsubsection{Optimized Buffering}
\label{sec:opt-buffering}
\ElectroPhotoAcc's throughput is limited by the rate at which data are input into the \PhotonicCore. 
While the latency and bandwidth of activation and weight SRAM arrays can be designed to match the throughput of the \PhotonicCore, the sizes of these arrays are limited.
If the activations and weights do not fit within these SRAM arrays, frequent DRAM accesses would be necessary.
These DRAM accesses are slower compared to SRAM accesses and can easily bottleneck the system performance. 
To avoid being bottlenecked by DRAM latency during runtime, we may want to limit the batch size for a given neural network. 
On the other hand, larger batch sizes provide a better throughput.
We, therefore, propose an optimized buffering method, which maximizes the batch size stored in the activation SRAM without ever spilling back to the DRAM during runtime.
This method takes advantage of the empty space in the SRAM during inference and loads the inputs of the next batch from DRAM efficiently.

We describe this optimized buffering method as a convex optimization problem.
Let $\vec{x}_\text{c}=[x_\text{c}(t_0),\;  x_\text{c}(t_1),\; \dots ,\; x_\text{c}(t_\text{max})]$ be a vector representing the activation SRAM array usage while performing inference on a batch of activations over time.
Here $t_\text{i+1}=t_i+\Delta t$ where $\Delta t$ is some time interval chosen to ensure the optimization problem is tractable for the host CPU.
Similarly, $\vec{x}_\text{pcie}=[x_\text{pcie}(t_0), \; x_\text{pcie}(t_1),\; \dots ,\; x_\text{pcie}({t}_\text{max})]$ is a vector representing the activation SRAM usage of the data (next input batch) being transferred from DRAM into SRAM over time.
For a given $\vec{x}_\text{c}$, an optimal $\vec{x}_\text{pcie}$ data transfer schedule can be obtained by solving the following optimization problem:

\begin{tabular}{l l}
$ \text{Maximize:} $& $\sum\limits_{t=t_0}^{t_\text{max}} {x_\text{pcie}(t)}$\\
\vspace{0.02in}
$\text{Subject to:}$ & $0\leq x_\text c(t)+x_\text{pcie}(t) \leq  x_\text{max};$\\
$\ $ & $x_\text{pcie}\geq 0$; $x_\text{pcie}(t_{-1})=0$; $x_\text{pcie}({t_\text{max}})=x_\text{input}$;\\
$ \ $ & $0\leq \Delta x_\text{pcie}(t)\leq \text {Max. PCI-e bandwidth}$\\
\vspace{0.05in}
\end{tabular}

The constraints in the optimization problem can be understood as: the total SRAM usage \begin {enumerate*} [label=(\arabic*\upshape)]
 \item should be less than the given SRAM size ($x_\text{max}$), \item should not be negative at any time, and \item should start from zero; \item the total amount of data transferred will be equal to the input size of the next batch, and \item the data transfer rate should be slower than the maximum PCI-e bandwidth.   
\end {enumerate*}
The objective function is to maximize the area under the curve of memory usage of the transferred data for the next batch.
Maximizing this area guarantees transferring the data \emph{as soon as possible} under the constraint of a maximum PCI-e bandwidth. 
If the program fails to return a schedule $x_\text{pcie}(t)$ that meets the specified constraints for a given batch size and maximum PCI-e bandwidth, a smaller batch size or a larger bandwidth (if it is available on the hardware) should be chosen.
We use the above optimization program to find the \emph{largest} batch size which ensures that the memory usage from storing activations of the current batch and the next batch never exceeds the SRAM size.
As such, we ensure that all DRAM data transfer for the next batch of inputs can happen simultaneously with the inference of the current batch.
The optimized schedule is computed only once by the host CPU before runtime.
\subsubsection{Parallelism}
\label{sec:parallelism}
\ElectroPhotoAcc can be scaled up to include multiple photo-cores.
We offer two parallelization strategies for distributing the workload among multiple photo-cores: data parallelism and tile parallelism. 
Data parallelism aims to accelerate MVMs by copying the same weights to all photo-cores.
Each \PhotonicCore performs the same operations on different inputs in a batch.
Tile parallelism is a finer granularity model parallelism that distributes different tiles of a weight matrix across multiple photo-cores.
Unlike data parallelism, all inputs in one batch are sent to all \PhotonicCores.

ADEPT can also use WDM-based parallelism.
WDM uses multiple wavelengths for encoding different input vectors at once similar to data parallelism. 
The scheme requires multiplexing and demultiplexing circuits that can be constructed from microring resonators~\cite{rings-bogaerts} or cascaded unbalanced MZIs~\cite{MZI-demultiplexer}.
WDM parallelism is synonymous to data parallelism in terms of throughput, but the same MZI array and weight DACs can be used by all inputs encoded in the wavelengths.

\begin{figure}[t]
   \centering
    \includegraphics[width=\linewidth]{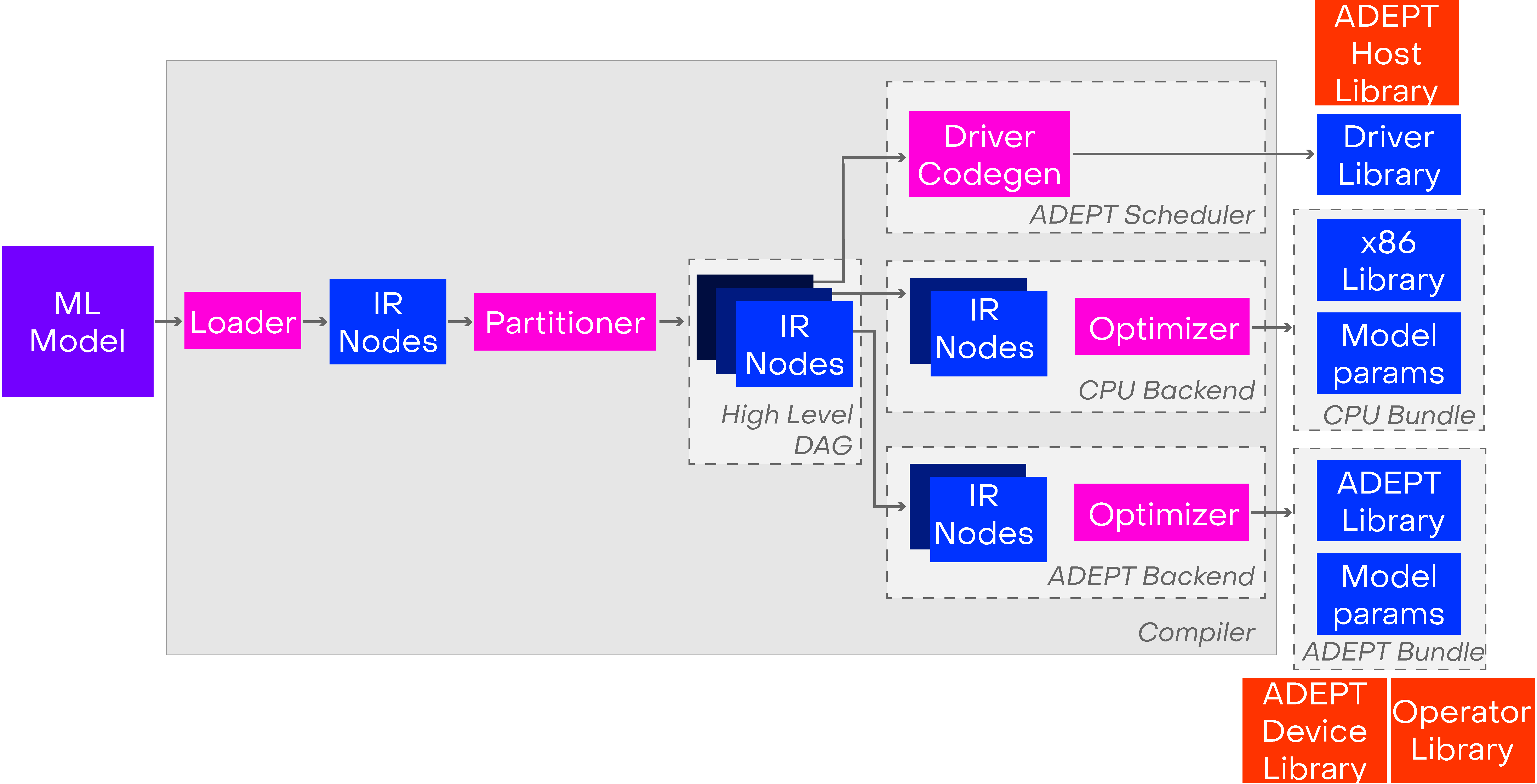}
    \caption{\textbf{Execution model.} Compilation process of an ML model for ADEPT.}
    \label{fig:compiler}
\end{figure}
\subsection{Execution Model}
\label{sec:compilation}

In this section, we describe the execution model for using ADEPT as part of the full-system. This process is summarized in Figure~\ref{fig:compiler}. 
Here, we take a DNN model and compile it on the host CPU to generate a program in the form of a graph on tensor types. 
We use ONNX models (exported from the common frameworks, such as Pytorch) and a loader to build a high-level program graph whose nodes are operations on higher dimensional array datatypes.
We create a directed acyclic graph (DAG) by using a cost-model based partitioner and annotate nodes based on whether the operations will be executed on a CPU or on the ADEPT device. 
We use an LLVM-based optimizer on the host CPU for code generation along with the optimizations. 
We then expand the operations annotated for execution on the ADEPT device into a stream of ADEPT instructions, and perform a scheduling pass to achieve overlap of GEMM operations and non-GEMM operations.
We use the annotated program graph to optimize the schedule and pipeline compute on the host CPU and the ADEPT device with communication between the two.
The generated code for these three partitions are linked with the corresponding libraries to produce two executable binaries: one for the host and one for \ElectroPhotoAcc.
It should be noted that the host CPU performs the compilation only once and then offloads the inference to \ElectroPhotoAcc.

\section{Evaluation Methodology}
\label{sec:eval-method}

In this section, we describe our evaluation approach when we compare ADEPT against SAs and state-of-the-art accelerators. We provide power, performance, and area analysis for both standalone GEMM cores, as well as for the full system.
For our evaluation, we choose three DNNs: ResNet-50\cite{resnet}, BERT-large~\cite{devlin2019bert}, and RNN-T \cite{rnnt}.
These three state-of-the-art networks---all part of the MLPerf inference data-center benchmarks~\cite{mlperf} in the \emph{offline} scenario---represent the diversity in layer types, sizes, and shapes that we observe in DNNs.
We combine architecture, circuit, and device level analyses to evaluate the full system. 

\subsection{Architecture-level Analysis} We used a mix of SCALE-Sim~\cite{scalesim} and RTL simulations for our architecture-level analysis.
SCALE-Sim is a simulator built for SA architectures. 
It takes the \SystolicArray configuration (i.e., array size and dataflow type) and the neural network configuration (i.e., layer sizes and batch size) as inputs, and calculates the number of cycles needed to execute the neural network. The simulator also generates traces for SRAM and DRAM reads/writes.
We modified SCALE-Sim to model the performance of the \PhotonicCore in \ElectroPhotoAcc.
The modifications were added on top of the existing WS dataflow in SCALE-Sim, which is similar with the WS approach of the photo-core.
These modifications include adding the latency for programming the weight tile into the MZI array, adding the latency for transferring the weights from the weight SRAM to the weight buffer and overlapping this data transfer latency with continuing GEMM operations. 

SCALE-Sim enables us to simulate our dataflow and directly compare the performance of the \PhotonicCore with that of \SystolicArrays. 
However, it only models GEMM operations.
To evaluate non-GEMM operations, we designed the \ElectronicAsic using SystemVerilog RTL.
We also incorporated the optimizations described in Section~\ref{sec:optimizations} in our evaluation.
For each DNN, we combined the timing results obtained from SCALE-Sim and RTL simulations to get the overall performance.

\subsection{Circuit/Device-level Analysis} 
For a realistic power, performance, and area comparison, we designed the \ElectronicAsic units and \SystolicArrays at RTL level and then synthesized them using Cadence Genus~\cite{genus} with a standard cell library designed in the GF22FDX technology node~\cite{22fdx}.
The SRAM arrays were generated using an SRAM compiler for GF22FDX.

To minimize impact of slow DRAM transfers on performance, prior works have used large on-chip memory arrays~\cite{kim2021bts,cerebras}.
We follow the same strategy.
However, it is challenging to have a single large SRAM array with low access latency. 
So, instead, we use multiple small sized SRAM sub-arrays to build larger memory arrays. 
The SRAM sub-arrays were designed to have 64 KB capacity with $\sim 1$ ns access latency. 
For higher clock frequencies ($f_c>1$ GHz), we read from multiple arrays, each offset by $1/f_c$ ns with its neighbor.
In total, we use 300 MB weight SRAM and 100 MB activation SRAM.
We acknowledge that not all our buses connecting the SRAM arrays to the photo-core will have the same latency. For a 700 mm$^2$ (reported in Section~\ref{sec:system-comparison}) chip size, the latency is calculated as $\sim$1.2~ns (maximum 12 cycles for the 10~GHz system) for the longest distance to travel (from one corner to the diagonally opposite corner)~\cite{chip-comm-joshi}. 
The throughput of the SRAM accesses is matched with the system clock by operating each SRAM sub-array at 833 MHz, but reading from the different SRAM sub-arrays every 100 ps.

The \PhotonicCore is powered by a laser. 
We calculated the required laser power per channel $P$ analytically by considering (1) the laser wall-plug efficiency, (2) the losses of the various optical devices, and (3) the SNR needed for an 8-bit output,
as follows:
\begin{equation}
\label{eq:laser_power}
    P = \frac{(\kappa \; \text{SNR}_{\text{shot}})^2  \cdot (q \; \Delta f/4)}{\eta_{\text{det}} \cdot \eta_{\text{array}} \cdot \eta_\text{mod} \cdot \eta_{\text{cpl}}\cdot \eta_{\text{laser}} },
\end{equation}
where $\text{SNR}_{\text{shot}} $ is the SNR assuming shot noise only and $\kappa$ (assumed to be $\approx 3$) accounts for noise contributions (e.g., thermal noise and transistor noise) other than the shot noise. The overall $\text{SNR} = \kappa \; \text{SNR}_{\text{shot}} = 2^{b_{\tout}}$ with $b_{\tout}$ being the bit precision of the output ADC. Here, $q$ is the elementary charge, and $\Delta f$ is the bandwidth of the coherent detector (related to the clock frequency). The $\eta$'s account for the transmissivity from the laser to the detectors. $\eta_{\text{mod}}$ is the transmissivity of the modulator ( $\approx1.2$~dB loss~\cite{Akiyama:12}), $\eta_{\text{array}}$ is the transmissivity of the MZI array ( $\approx 0.04$~dB loss per MZI~\cite{noems} and each signal passes through $2m+1$ MZIs), $\eta_{\text{cpl}}$ is the fiber laser-to-chip coupling efficiency ($\approx 2$~dB loss), $\eta_{\text{det}}$ is the efficiency of the photodetectors ($\approx 80\%$~\cite{photo-det}), and $\eta_{\text{laser}}$ is the wall-plug efficiency of the laser ($\approx 20\%$~\cite{soton356442}).
All the photonic devices in the photo-core are simulated using Lumerical Maxwell-Equations solver FDTD and circuit-level simulator INTERCONNECT~\cite{lumerical}. They have also been fabricated in the GF90WG SiPh process and are characterized at multiple-wafer-scale with the FormFactor CM300 wafer tester.

The necessary bit precisions for the inputs and the weights are 10 bits and 12 bits, respectively, to guarantee the 8-bit-precise outputs read by the ADCs (See Section~\ref{sec:precision}).
Due to the lack of publicly available DAC prototypes in GF22FDX with our desired precision, for our analysis, we used a 14-bit DAC~\cite{14bit-dac} designed with 28 nm CMOS technology with a 10 GS/s sampling rate and 177 mW power consumption. Note that the power consumption of 10-bit and 12-bit DACs will be less than a 14-bit DAC. Therefore, we scaled the power numbers as follows:
A widely accepted figure of merit (FoM) for the performance of DACs is $\text{FoM} = 2^B \cdot \left. f_s \right\rvert_{6(B-1)} /P_{\text{DAC}}$.
Here, $B$ is the bit precision of the DAC, $\left. f_s \right\rvert_{6(B-1)}$ is the output signal frequency where the spurious free dynamic range has dropped with 6~dB (= 1~bit) in comparison with the expected results ($\approx 6B$), and $P_{\text{DAC}}$ is the power consumption of the whole DAC~\cite{Li_2020}.
In essence, the power consumption of a DAC---with the same FoM---is proportional to $2^B$.
Therefore, a 12-bit DAC (for the weights) with the same FoM will consume $2^2=4$ times less power than a 14-bit DAC.
Similarly, a 10-bit DAC (for the inputs) with the same FoM will consume $2^4=16$ times less power than a 14-bit DAC. 
The 10-bit input and 12-bit weight DACs will then consume 11.06 mW and 44.25 mW, respectively.
Similar to DACs, we use 10-bit ADCs in 28 nm technology at the output.
Within the 10 ns settling time constraint of the MZIs, a single 10~GS/s DAC can be used to program 100 weights into MZIs.
Therefore, instead of using $m^2$ DACs, we use $\lceil m^2/\zeta \rceil$ DACs for weights where $\zeta$ is equal to 100.
Each ADC has a 5 GS/s sampling rate and consume 29 mW~\cite{adc}.
The electronic-to-optical (E-O) and optical-to-electronic (O-E) conversion power is based on the total energy required to operate the modulator circuitry, which is $\sim 20$~fJ/bit, and the detector circuitry, which is $\sim 297$~fJ/bit~\cite{Sun2015SinglechipMT}.
Each DRAM access is assumed to be 20~pJ/bit~\cite{horowitz-energy}.
The die-to-die interconnect between the photonic and electronic chiplets consumes 0.3 pJ/bit~\cite{7552548}
\section{Evaluation Results}

Our evaluation focuses on answering two questions: (1) how do we build a complete electro-photonic accelerator architecture that
is not bottlenecked by the slower electronics? (2) how much are the electro-photonic accelerator systems
better than purely electronic accelerator systems, when we consider the system as a whole, i.e., accelerator
+ memory + host processor + communication, running practical real-world applications?

We preferred SAs for comparison as they are commonly used for DNN acceleration.
SAs provide high throughput and efficiency, and have a similar dataflow as the photo-core. 
In Section~\ref{sec:pc-vs-sa}, to set the stage, we first provide a detailed comparison of standalone photo-core against electronic SAs. 
This comparison helps us determine the ADEPT design that we should use for exploring different architecture optimizations as well as for performing full-system analysis.
In Section~\ref{sec:optimizations-eval}, we evaluate the impact of optimizations we introduced in~\ref{sec:optimizations} and in Section~\ref{sec:parallelism-eval}, we analyze the different parallelism methodologies. 
While these first three sections answers the first question, Section~\ref{sec:system-comparison} answers the second question by comparing the complete ADEPT-based system where all the components and optimizations are taken into account against a similar system where photo-core is replaced with a same-sized SA. 
Lastly, in Section~\ref{sec:sota-acc}, for completeness, we provide a comparison of ADEPT against state-of-the-art electronic and photonic accelerators. 

\begin{figure}[t]
  \centering
  \includegraphics[width=\linewidth]{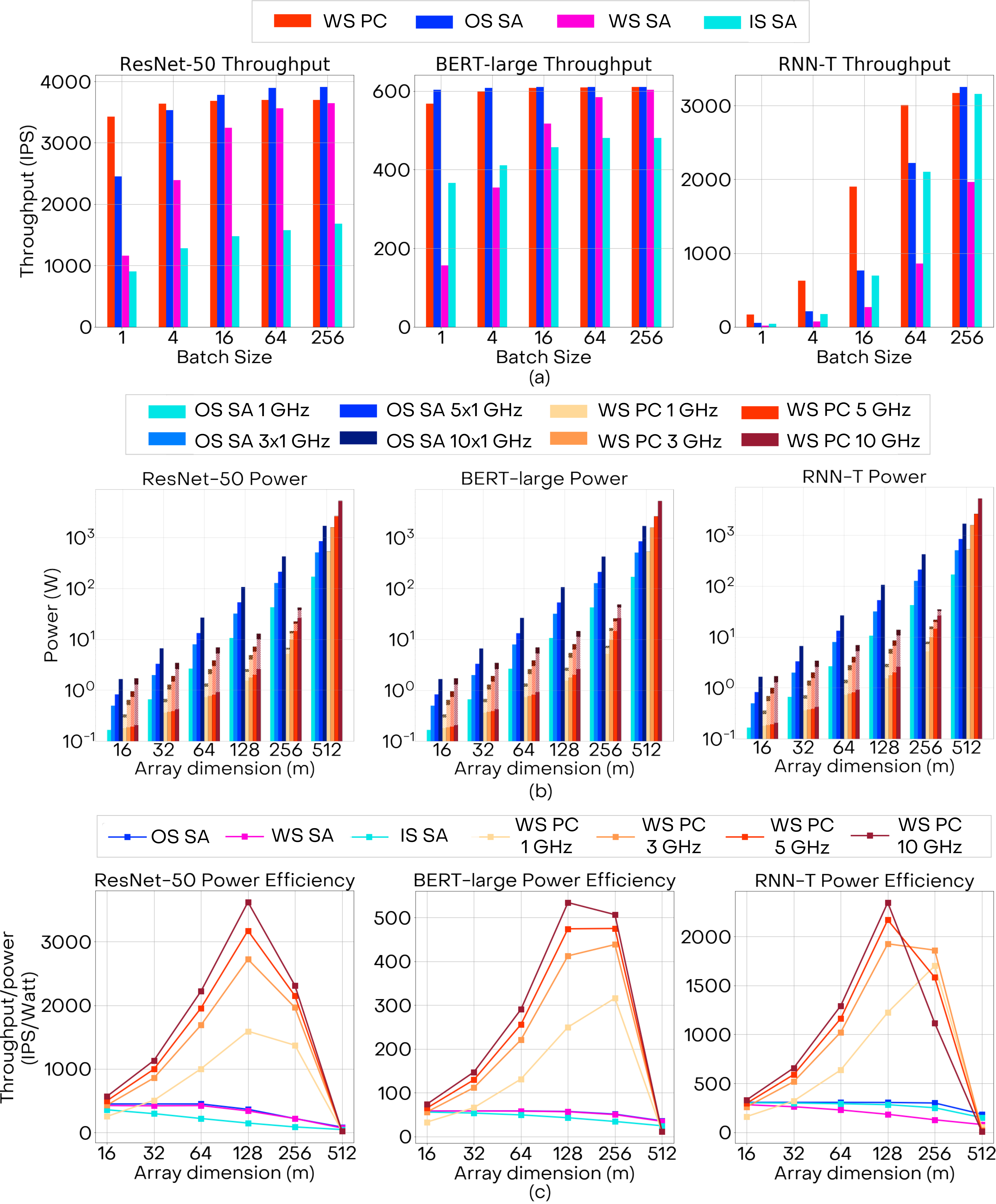}
  \caption{\textbf{Photo-core vs SA comparison in terms of throughput (IPS), power (W) and power efficiency (IPS/W)} (a) Throughput vs. batch size of $128 \times 128$ \PhotonicCore (PC) and \SystolicArrays with OS, WS, and IS dataflows at 1 GHz clock. 
  (b) Power consumption of the OS \SystolicArray (1, 3$\times$1, 5$\times$1, and 10$\times$1~GHz clock) and the WS \PhotonicCore (1, 3, 5, and 10~GHz clock), for different array sizes. For the \PhotonicCore we report power in laser (solid color), ADCs/DACs (white diagonal pattern) and E-O/O-E conversion (black diagonal pattern).
  (c) Power efficiency (in IPS/W) of OS, WS, IS \SystolicArrays (1, 3$\times$1, 5$\times$1, and 10$\times$1~GHz clock) and WS \PhotonicCore (1, 3, 5, and 10~GHz clock) for different array sizes.
  Here $f_c \times 1$ GHz SA indicates that we use $f_c$ SAs in parallel, each operating at 1 GHz.}
  
\label{fig:mvp-analysis}
\end{figure}

\subsection{Photo-core vs. SAs}
\label{sec:pc-vs-sa}
The \PhotonicCore utilizes light, which oscillates at hundreds of terahertz, and so it has a significant bandwidth advantage over the electronic \SystolicArrays. 
The bandwidth in the photo-core is typically limited by the sampling rate of data converters (considered up to 10 GHz in this work), while \SystolicArrays are constrained due to parasitic resistance, capacitance, and inductance.
In fact, in case of \SystolicArrays, Cadence Genus with GF22FDX failed to meet the timing requirements for 2~GHz and above.
Therefore, we used parallelism instead to effectively operate the \SystolicArray at higher frequencies.
For example, to operate a \SystolicArray at 10~GHz, we used ten 1~GHz \SystolicArrays whose clock cycles are offset by 100~ps.
The latency of this parallelized \SystolicArray will still be 1~ns, but its throughput will be synonymous to a single \SystolicArray operating at 10~GHz.
For this analysis, we assume both photo-core and SA are isolated from the system, weights and inputs have been loaded and are available in the SRAM with reads/writes fast enough to keep up with the requirements of both arrays.
To provide SAs a strong baseline, we considered OS, WS, and IS for \SystolicArrays as dataflow can have a significant impact on SA's performance.
Fig.~\ref{fig:mvp-analysis}(a) shows the throughput we can achieve when using the three dataflows for \SystolicArrays for three different benchmarks.
We observe that OS performs better than WS and IS for \SystolicArrays.
This is because of the high latency of loading data into the \SystolicArray between tiles for WS and IS dataflows.
Therefore, from here onwards, we use OS for \SystolicArray in the rest of the comparison.

\subsubsection{Throughput}

For the throughput comparison, we use a single $128\times128$ array (the choice of size is justified later in this section) for both the \PhotonicCore and the \SystolicArray.
Fig.~\ref{fig:mvp-analysis}(a) shows the comparison between the performance of the \PhotonicCore (when using WS dataflow) and that of the \SystolicArray operating at 1~GHz clock frequency for different batch sizes for three different networks---ResNet-50, BERT-large, and RNN-T (one plot per network).
In photo-core, by pipelining the weight transfer from weight SRAM into the weight buffer with GEMM operations, we reduce the latency of loading the weights down to 10 ns, the minimum required by MZIs (see Section~\ref{sec:photonic-unit} and ~\ref{sec:memory}).
In general, photo-core's WS dataflow is more advantageous compared to the OS SA when the weight matrices are large and the input matrices are small (e.g., RNN-T with small batch sizes) because each weight tile needs to be loaded only once.

\textbf{Throughput vs. Batch Size: }
From Fig~\ref{fig:mvp-analysis}(a) we can see that as the batch size increases, throughput (and correspondingly utilization) of the arrays increases and eventually saturates.
Among the three DNNs, we observe that the throughput saturates for ResNet-50 and BERT-large more quickly than RNN-T.
This is because the small input matrices in RNN-T means that fewer number of vectors are multiplied with the same tile. 
Thus, the utilization and throughput continue to significantly increase until we have larger batch sizes for RNN-T.
In addition, as the batch size increases, latency in between tiles becomes less important because more time is spent on performing MVM operations in each tile.  

\textbf{Throughput vs. Operating Frequency:}
One way to increase the throughput of any computing device is to increase the clock frequency.
We therefore attempt to increase the clock frequency of the \PhotonicCore and the \SystolicArrays (from 1~GHz to 3~GHz, 5~GHz, and 10~GHz).
The throughput of the \SystolicArrays increases linearly with the clock frequency.
The rate of MVM operations in the photo-core also increases linearly with the clock frequency. 
However, a fixed 10~ns period is necessary for programming the MZI array and is independent from the clock frequency. 
Therefore, the increase in the throughput of the photo-core is sub-linear. 

\subsubsection{Power Consumption}

Fig.~\ref{fig:mvp-analysis}(b) compares the average power consumed by the WS \PhotonicCore (laser, ADC/DAC, and E-O/O-E conversion) and the OS \SystolicArray of different sizes. 
For this analysis, we use a batch size of 256 to ensure that the throughput is nearly saturated for all networks.
Overall, the \PhotonicCore's power consumption is smaller than the \SystolicArray counterpart up to an array size of $256 \times 256$.

For the \SystolicArrays, the power consumption increases linearly with the number of PEs (quadratically with the array size). 
For the \PhotonicCore, the laser power increases exponentially with the depth $m$ of the array, due to optical loss. 
As a result, it can be seen in Fig.~\ref{fig:mvp-analysis}(b) that laser power dominates for larger array sizes. 
For an $m\times m$ \PhotonicCore, we need $m$ DACs and $m$ E-O conversion circuits for the input vector, and $m$ ADCs and $m$ O-E conversion circuits for the output vector. 
These input/output DACs/ADCs perform a conversion each cycle. 
Additionally, we need DACs for programming the $m\times m$ weight matrix. These DACs for programming the weights into the MZIs are not used each cycle.
The weights are programmed into the MZI once for each tile, and the DACs are not used until all MVMs for the corresponding tile are finished. 
The average power consumption of DACs/ADCs increases as the array size increases because the latency drops.
Effectively, the same number of conversions are performed within a shorter duration of time. 

\subsubsection{Power Efficiency}
Fig. \ref{fig:mvp-analysis}(c) shows the power efficiency (IPS/W) of electronic \SystolicArrays and \PhotonicCores for different array sizes and frequencies. 
We observe that, for the \PhotonicCore, $128 \times 128$ is the most power-efficient array size for all three networks and all four clock frequencies. 
This can be explained by the fact that beyond a certain size, the laser power starts dominating the power consumption of the \PhotonicCore.
Additionally, beyond a certain array size, the utilization decreases and so the throughput saturates.
Therefore, due to the exponentially increasing laser power and saturating throughput, we observe a drop in the power efficiency beyond an array size of $128 \times 128$.
For \SystolicArrays, the power increases quadratically with the array dimension $m$. 
However, because the throughput increases less than quadratically with $m$, the power efficiency decreases as the array size increases.

Across different frequencies and array dimensions, we observe that \PhotonicCore can provide up to \textbf{9.87}$\times$, \textbf{9.32}$\times$, and \textbf{7.69}$\times$ better power efficiency than OS \SystolicArray for ResNet-50, BERT-large, and RNN-T, respectively, when only GEMM operations are considered. 
\textbf{Overall, we observe that for the same clock frequency, while the throughput is comparable, photo-core provides a better power efficiency than the best performing SA.} As we show that $128 \times 128$ is the most power-efficient array size for the photo-core, we will use this array size for the further evaluations.

\subsection{Optimizations}
\label{sec:optimizations-eval}

As discussed in Section~\ref{sec:optimizations}, non-GEMM operations and data transfers introduce latency and energy overhead, and are important in system evaluation.
In this section, we quantify these overheads and show the impact of the optimizations we apply on performance of ADEPT for different types of DNNs.

\subsubsection{Pipelining}

\begin{figure}[t]
\centering
\includegraphics[width=\linewidth]{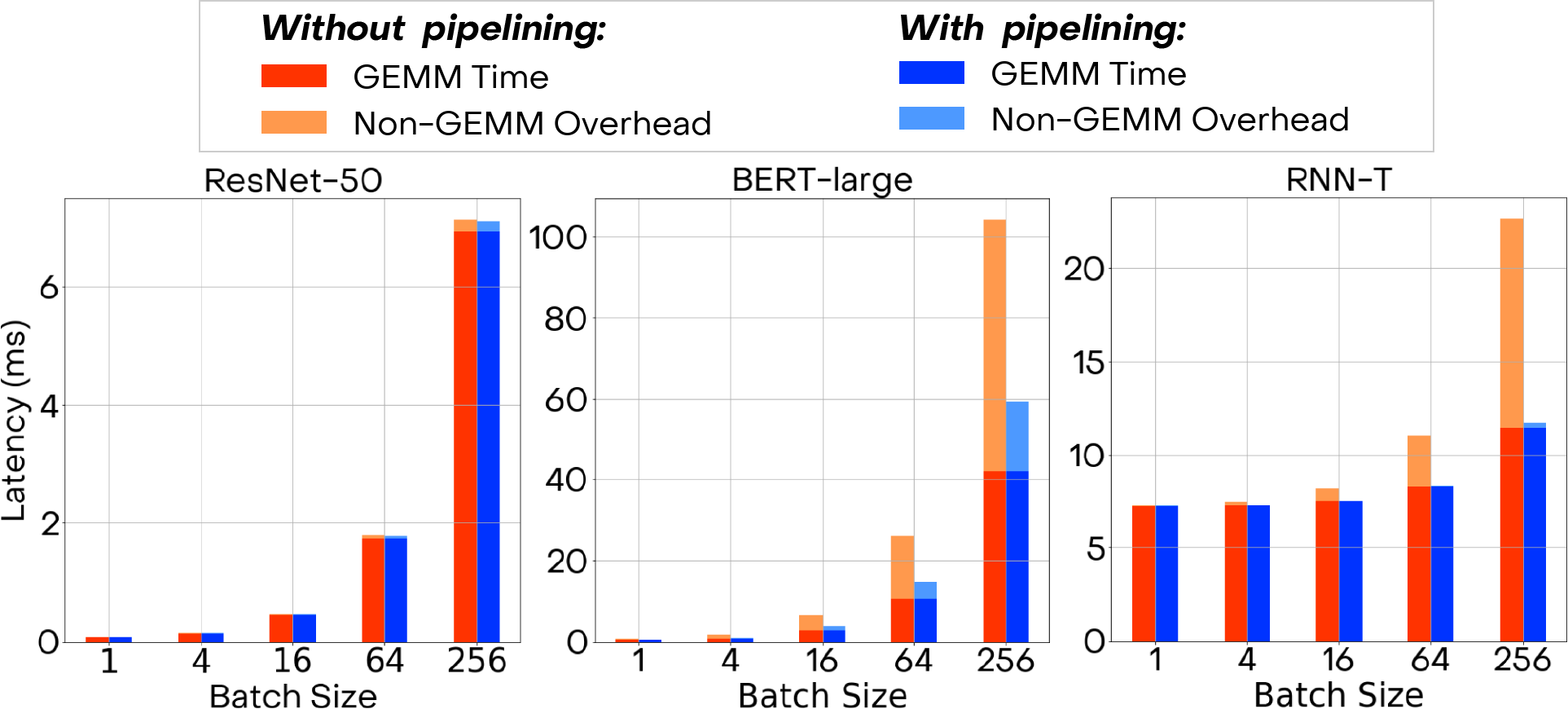}
  \caption{\textbf{Pipelining evaluation.} Latency of \ElectroPhotoAcc with a $128 \times 128$ \PhotonicCore operating at 10 GHz clock with and without pipelining the GEMM and non-GEMM operations. Here the latency is for one batch of inputs for three networks. The results are calculated for varying batch sizes.}
  \label{fig:dig-op}
 \end{figure}

Fig. \ref{fig:dig-op} shows the impact of pipelining operations on the inference time of \ElectroPhotoAcc when running ResNet-50, BERT-large and RNN-T. 
For ResNet-50, the max-pool, average-pool, ReLU activations and softmax layers; for BERT-large, the layer norm, GELU and softmax operations; and for RNN-T, the element-wise addition and multiplication, sigmoid, and tanh operations (within an LSTM layer) are computed in the \ElectronicAsic.
The non-GEMM operations comprise a small percentage of the networks' operations, but they can lead to a large overhead if not pipelined carefully. 
When pipelined, the non-GEMM operations and the GEMM operations can be performed in parallel.

In Fig. \ref{fig:dig-op}, we can see that ResNet-50 has the least amount of overhead due to non-GEMM operations. 
With batch normalizations folded, ReLU becomes the most frequent non-GEMM operation, which can be effectively overlapped with the GEMM operations. 
In BERT-large and RNN-T, the division and exponential operations in GELU, softmax, sigmoid, and tanh increase the number of cycles spent in the \ElectronicAsic.
As batch size increases, GEMM operations are performed more efficiently because more input vectors are multiplied with the same tile---weights are re-used more frequently.
On the other hand, the cycles spent on non-GEMM operations increase linearly with batch size. 
Effectively, we observe a larger increase in the time spent on the non-GEMM operations than the increase in time spent on the GEMM operations with increasing batch size. 
As a result, a smaller portion of the non-GEMM operations can be overlapped with the GEMM operations.
We observe a reduction in latency of up to 5.73$\%$ in ResNet-50, 43.03$\%$ in BERT-large, and 48.22$\%$ in RNN-T when we pipeline the non-GEMM and GEMM operations. 

\subsubsection{Optimized Buffering}
\label{sec:opt-buffering-eval}
\begin{figure}[t]
\centering
\includegraphics[width=\linewidth]{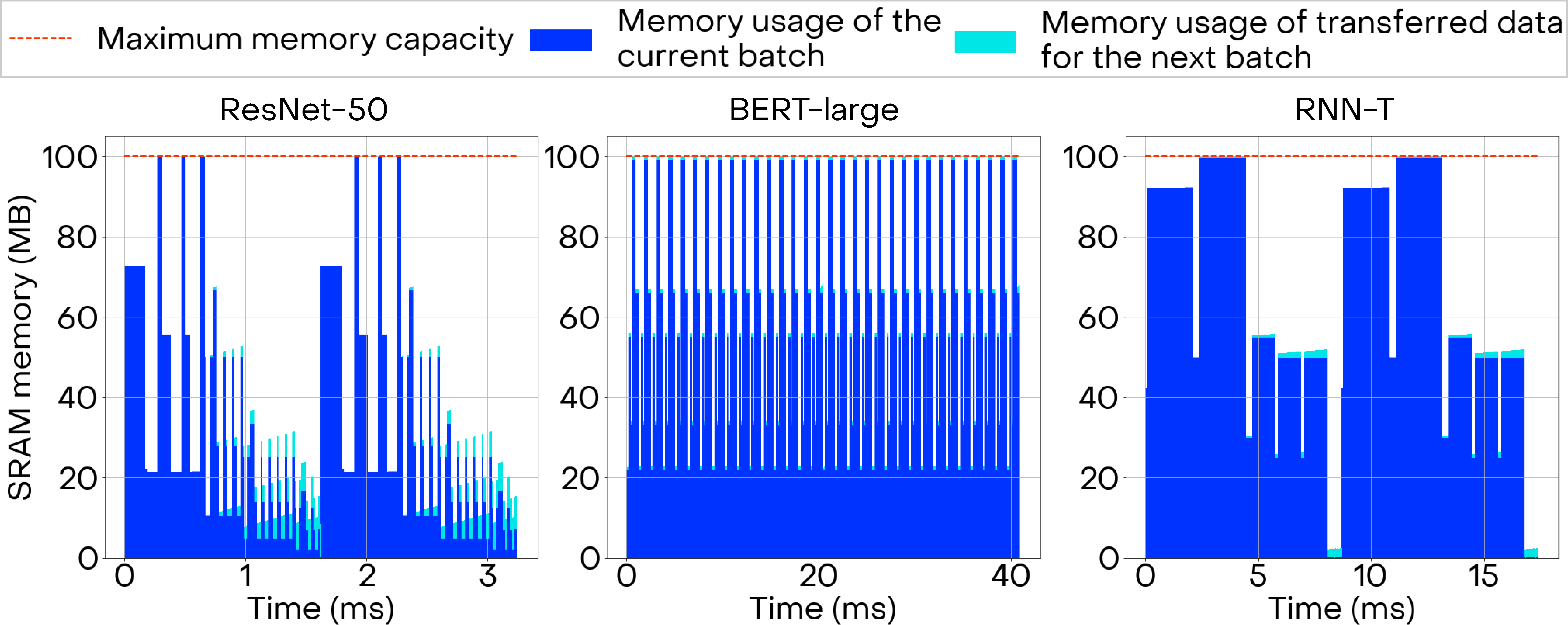}
  \caption{\textbf{Activation SRAM usage for computing on the current batch of inputs along with data transfer for the next batch of inputs within \ElectroPhotoAcc.} Both input and output activations for the current batch must be stored in the activation SRAM (dark blue) while the input data are transferred for the next batch (light blue). A $128 \times 128$ \PhotonicCore  at 10 GHz clock is used with batch sizes of 58, 88, and 50 for ResNet-50, BERT-large, and RNN-T, respectively to fully use the 100 MB activation SRAM capacity.}
  \label{fig:dma}
\end{figure}

Up until now, we used a large batch size of 256 to evaluate the saturated throughput of both ADEPT and \SystolicArrays.
However, given that the SRAM arrays have limited sizes, an inference with batch size of 256 may not fit within the activation SRAM. 

For this analysis, we choose a 100 MB activation SRAM and a 300 MB weight SRAM to ensure that the weights of all the three networks can comfortably fit within \ElectroPhotoAcc.
Fig. \ref{fig:dma} shows the usage of the activation SRAM array for the current batch and the next batch when using our optimized DRAM access mechanism (See Section~\ref{sec:opt-buffering}).
We limit the batch size to the maximum value where inference on the entire batch can be completed without any DRAM transfers (58, 88, and 50 for ResNet-50, BERT-large, and RNN-T, respectively).
The activation SRAM stores the inputs and outputs of all GEMM and non-GEMM operations over time. 
If the GEMM and non-GEMM operations are running at the same time (pipelined), the memory usage includes both of the operations' activation data. 
Fig. \ref{fig:dma} shows that the networks do not use the whole SRAM array throughout the inference. 
This creates an opportunity to transfer the inputs for the next batch. 

We compare the performance of our optimized buffering technique against double buffering~\cite{4536316}: a common method for minimizing the impact of data transfer latency. 
In double buffering, one half of the memory is used for the current inference while the other half is used for transferring the inputs for the next inference.
As a result, the maximum batch sizes of this scheme, for the three networks, are half of those of the optimized buffering scheme.
For ResNet-50 and BERT-large, optimized buffering technique increases the throughput only by $1.3 \%$ and $0.4 \%$ compared to double buffering.
This is because these two networks have already high utilization in the photo-core and their throughputs are saturated for the considered batch sizes.
Remarkably, however, optimized data transfer increases the throughput of RNN-T by $89.7 \%$ over double buffering.

\subsubsection{Impact of Optimizations}

\begin{figure}[t]
\centering
\includegraphics[width=\linewidth]{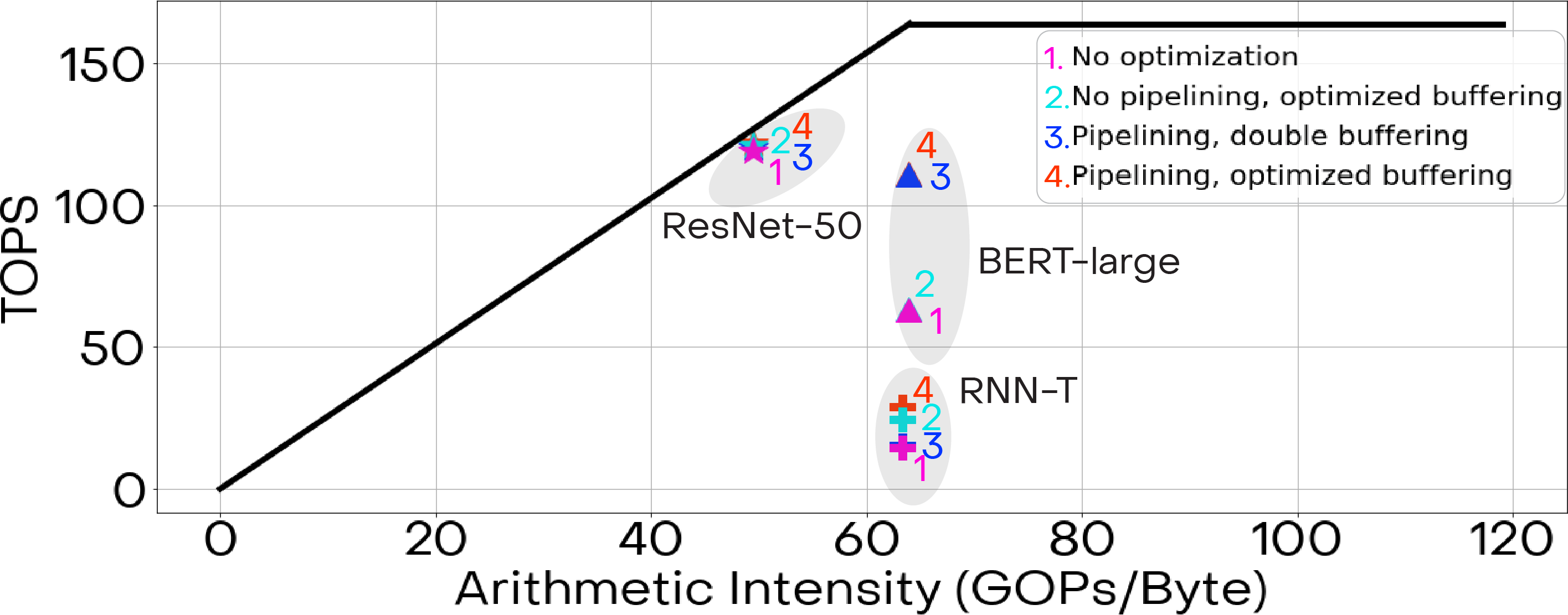}
  \caption{\textbf{Impact of optimizations. }Roofline plot showing the effect of optimizations on \ElectroPhotoAcc with a single $128\times128$ photo-core. 
  The arithmetic intensity is calculated using MAC operations over activation SRAM reads/writes. }
  \label{fig:roofline}
\end{figure}

Fig.~\ref{fig:roofline} summarizes the impact of the two optimizations---pipelining and optimized DRAM buffering, on \ElectroPhotoAcc at a system level. The roofline is the peak throughput of the photo-core, and the memory ceiling is derived from the bandwidth of the activation SRAM.
The baseline (no optimization) refers to the case without any pipelining and with double buffering. 

Comparing the three networks, ResNet-50 has a smaller arithmetic intensity (AI) and is memory-bound.
We see that the performance of ResNet-50 without the optimizations is close to the roofline; thus, further optimizations only marginally improve the performance.
BERT-large significantly benefits from pipelining with a 1.76 $\times$ better throughput because of the frequent non-GEMM operations. 
In contrast, using the optimized DRAM buffering, which enables us to use larger batch sizes compared to double buffering, does not help because of the already saturated utilization of the photo-core for small batch sizes.
RNN-T has a lower utilization compared to the other two networks.
The utilization is mainly limited by the recurrent nature of the network, which requires frequent change of weight tiles and the frequent non-GEMM operations in the LSTM layers.
Therefore, increasing batch size by using the optimized DRAM buffering increases the performance significantly---by 1.92 $\times$ and pipelining improves the throughput for RNN-T by 1.83 $\times$. 

The analysis presented in this section highlights the importance of taking non-GEMM operations and memory limitations into account and using different types of DNNs for evaluation. 
\textbf{The non-GEMM operations and memory limitations limit the throughput of photo-core, but it is possible to go around these limitations and improve the performance by using the right optimizations such as pipelining and efficiently buffering the data. }

\subsection{Parallelism}
\label{sec:parallelism-eval}
\begin{figure}[t]
\centering
\includegraphics[width=\linewidth]{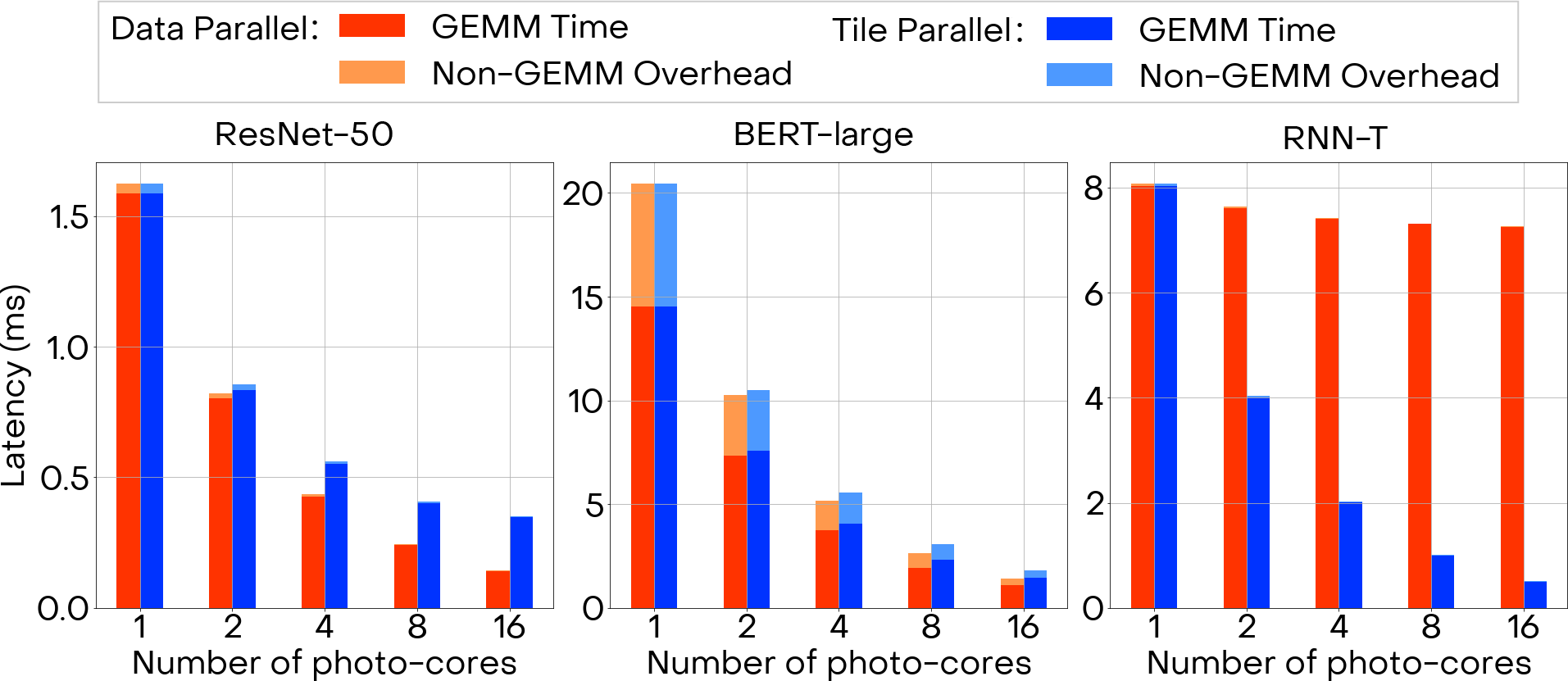}
  \caption{\textbf{Parallelism evaluation. }Latency of ADEPT ($128 \times 128$ photo-core at 10 GHz clock) when executing the three neural networks with different \PhotonicCore counts using data parallelism and tile parallelism.}
  \label{fig:multicore}
\end{figure} 

We consider three types of parallelism: data parallelism, tile parallelism, and WDM parallelism (See Section \ref{sec:parallelism}). 
Fig. \ref{fig:multicore} shows how the latency scales with increasing number of \PhotonicCore counts for both data and tile parallelism.
We use the batch sizes previously considered (see Section~\ref{sec:opt-buffering-eval}), i.e., 58, 88, and 50 for ResNet-50, BERT-large, and RNN-T, respectively. 
We keep these values constant as we increase the number of \PhotonicCores.

Data parallelism provides an almost linear decrease in inference latency with increasing photo-core count when the number of input vectors within a batch is large enough to be shared among the \PhotonicCores. 
The latency is dominated by MVM operations for large inputs sizes, and so as the number of photo-cores increases, the throughput proportionally increases.
We observe this in ResNet-50 and BERT-large where the input matrices are large enough to be spread among the photo-cores and we can maintain high utilization. 
In contrast, when the number of input vectors per core decreases, the reduction in latency saturates due to the decrease in the utilization of the \PhotonicCores. 
We observe this in RNN-T.
Data parallelism provides 11.30$\times$, 14.47$\times$ and 1.11$\times$ lower latency for ResNet-50, BERT-large, and RNN-T when we increase the \PhotonicCore count from 1 to 16.

The advantage of tile parallelism is limited by the number of tiles in a GEMM layer. 
The networks with larger weight matrices (i.e., BERT-large and RNN-T) better exploit this parallelism.
Tile parallelism provides 11.24$\times$, 16.0$\times$ and 4.62$\times$ lower latency for BERT-large, RNN-T and ResNet-50, respectively, when the \PhotonicCore count increases from 1 to 16.

Multiple \PhotonicCores means a linear increase in the area and the power consumption for the analog photonic computing unit.
WDM provides an opportunity to reduce this area increase.
WDM allows the input vectors to be mapped across the different wavelengths that are routed to same \PhotonicCore.
Therefore, WDM offers the same throughput as data parallelism without using multiple copies of the MZI array and weight DACs.
When we compare data parallelism with $n$ \PhotonicCores against a single \PhotonicCore leveraging $n$ wavelengths in WDM, the \PhotonicCore with WDM uses $(n-1) m^2$ fewer MZIs and $(n-1) m^2/\zeta$ fewer weight DACs. 
As an example, for $n=4$ wavelengths, on average across ResNet-50, BERT-large, and RNN-T, using WDM results in 1.41$\times$ better power-area efficiency (IPS/W$\cdot \text{mm}^2$) compared to using 4 parallel \PhotonicCores in a data parallel manner. 
This increase of power-area efficiency goes up to 1.53$\times$ for $n=8$ and 1.87$\times$ for $n=16$.
\textbf{Broadly, our analysis shows that different DNNs can benefit from different parallelization strategies and WDM can provide a better area efficiency compared to using multiple photo-cores.}

\subsection{System-level Comparison}
\label{sec:system-comparison}
\begin{figure*}[t]
\centering
\includegraphics[width=\textwidth]{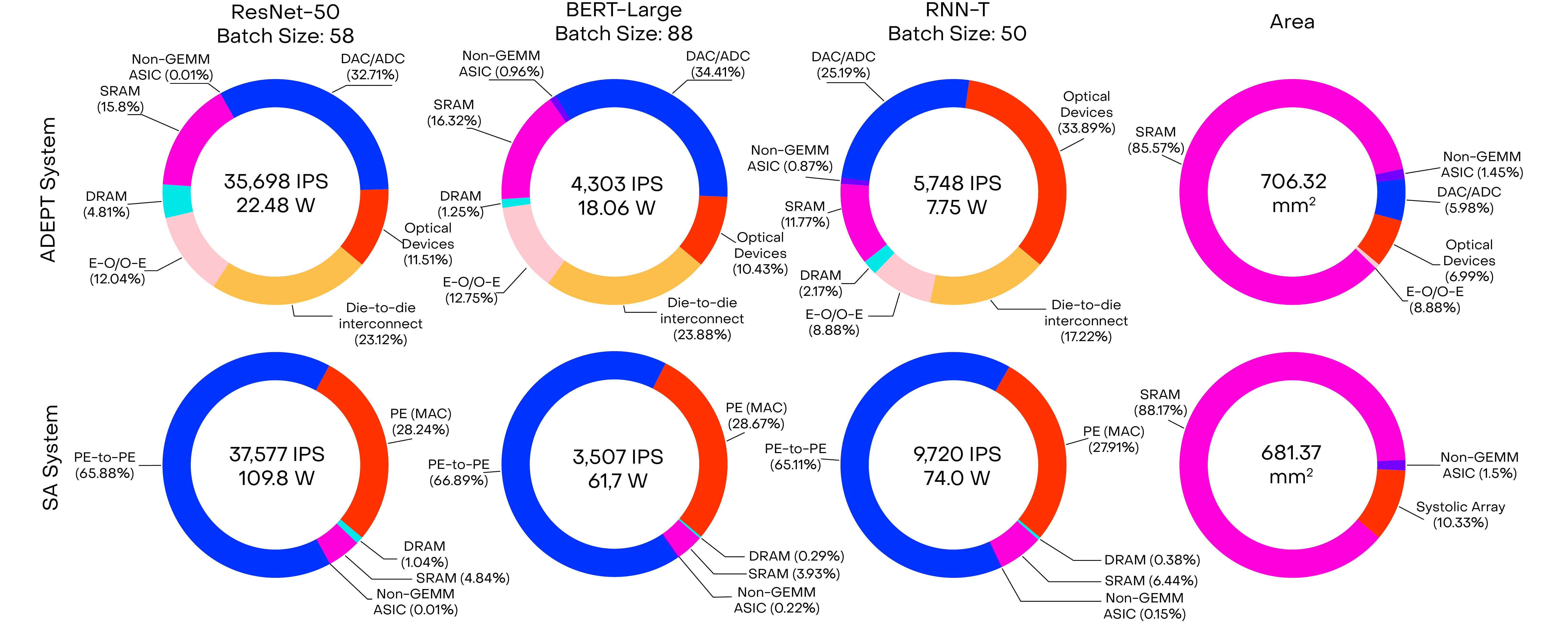}
  \caption{\textbf{\textit{Average total (static and dynamic) power distribution and area distribution of \ElectroPhotoAcc ($128 \times 128$, 10 GHz \PhotonicCore) and the {\SystolicArray} system ($128 \times 128$, 10$\times$1 GHz array, OS dataflow).}}}
  \label{fig:sys-comparison}
\end{figure*}

In this section, to answer the main question of how much the real benefit in a complete system is, we include all the components of the system and the optimizations discussed in Section~\ref{sec:optimizations}, and provide a full system-level comparison between $128 \times 128$ WS \ElectroPhotoAcc and a $128 \times 128$ OS \SystolicArray (see Fig.~\ref{fig:sys-comparison}).

From Figure~\ref{fig:sys-comparison}, we can see that the optical devices in the photo-core (i.e., laser, MZIs, modulators) used for the GEMM computation take up only between 10-35$\%$ of the overall power consumption in the ADEPT system depending on the DNN model. The other components of the system (i.e., ADCs/DACs, O-E/E-O conversions, die-to-die communication and SRAM) consume significant power---which proves the necessity of the system-level evaluation. 
For the SA, data transfer between the register files of the processing elements (PEs) dominate the power consumption of the SA system.
We observe that SRAM dominates the area distribution for both electronic \SystolicArrays and \ElectroPhotoAcc for the chosen configuration.

In ADEPT, the photo-core and the digital electronic ASIC are in different chiplets to take advantage of the technology nodes that provide the best performance for each individual electronic and photonic ICs. 
The two chiplets are 3D integrated through an interposer.
In the latter, the SA and the rest of the electronic components share the same chiplet.
The optimizations used for \ElectroPhotoAcc are also applied to the \SystolicArray system. 

Our analysis shows that a system with \ElectroPhotoAcc consumes 4.88$\times$ (109.8~W vs. 22.48~W), 3.42$\times$ (61.7~W vs. 18.06~W) and 9.55$\times$ (74.0~W vs. 7.75~W) less power for ResNet-50, BERT-large, and RNN-T, respectively. This translates to \textbf{4.89$\times$}, \textbf{3.24$\times$} and \textbf{9.06$\times$ } better power efficiency (IPS/W). Also, ADEPT  provides \textbf{4.5$\times$}, \textbf{2.97$\times$} and \textbf{8.34$\times$} better power-area efficiency (IPS/W$\cdot \text{mm}^2$) compared to a \SystolicArray.
\textbf{This shows us that although including the system components in evaluation decreases the performance of the standalone photo-core, we can still benefit from using photo-cores instead of SAs in a system. }

\subsection{Comparison Against DNN Accelerators} 
\label{sec:sota-acc}
For completeness, in this section, we compare the full ADEPT system against state-of-the-art electronic~\cite{eyeriss_v2, Eyeriss2017, unpu, tpuv3} and photonic~\cite{albireo, dnnara, holy-light} accelerators.

\subsubsection{Electronic Accelerators}

Besides the traditional \SystolicArrays, more flexible electronic accelerator architectures have been proposed and shown to perform more efficiently. 
Table~\ref{table:sota-elec-acc} compares \ElectroPhotoAcc against state-of-the-art electronic accelerators.
Much of the prior work focuses on AlexNet, and so we added ADEPT's results for AlexNet.
Broadly, for AlexNet and ResNet-50 inference, while is not the most area efficient, \ElectroPhotoAcc provides at least 6.8$\times$ higher IPS/W than other electronic accelerators.

\begin{table*}[h]
    \centering
    \caption{Comparison against state-of-the-art electronic and photonic accelerators.}
    \label{table:sota-elec-acc}
    \begin{tabular}{lccccccc}
        \toprule
        {} & \multicolumn{2}{c}{\textbf{ADEPT (This work)}} & \textbf{Eyeriss} \cite{Eyeriss2017} & \textbf{Eyeriss v2} \cite{eyeriss_v2}  & \textbf{UNPU} \cite{unpu} & \textbf{TPU v3} \cite{tpuv3}  \\
        \cmidrule(lr){2-3}
        \cmidrule(lr){4-4}
        \cmidrule(lr){5-5}
        \cmidrule(lr){6-6}
        \cmidrule(lr){7-7}
        \textbf{Tech Node }& \multicolumn{2}{c}{90 nm photonics + 22 nm CMOS} & 65 nm & 65 nm & 65 nm &  16 nm \\
        \textbf{Clock rate} & \multicolumn{2}{c}{10 GHz} & 200 MHz & 200 MHz & 200 MHz & 940 MHz \\
        \textbf{Benchmark} & AlexNet & ResNet-50 & AlexNet & AlexNet & AlexNet &  ResNet-50 \\
        \textbf{Batch size} & 192 & 58 & 4 & 1 & 15 & N/A \\
        \textbf{IPS} & 217, 201& 35,698 & 35 & 102 & 346 & 32,716 \\ 
        \textbf{IPS/W} & 7,476.78 &  1,587.99 & 124.80 & 174.80 & 1,097.50 & 18.18   \\ 
        \textbf{IPS/W/$\text{mm}^2$ }&  10.59 & 2.25 & 10.18 & N/A & 68.59 & 0.01  \\
        
\bottomrule
    \end{tabular}
\end{table*}

\begin{table*}[h]
    \centering
    \caption{Comparison against state-of-the-art photonic accelerators.}
    \label{table:sota-photonic-acc}
    \begin{tabular}{lcccccccc}
        \toprule
        {} & \multicolumn{4}{c}{\textbf{ADEPT (This work)}} & \textbf{Albireo-C} \cite{albireo} & \textbf{DNNARA} \cite{dnnara} &\textbf{HolyLight-A} \cite{holy-light}  \\
        \cmidrule(lr){2-5}
        \cmidrule(lr){6-6}
        \cmidrule(lr){7-7}
        \cmidrule(lr){8-8}
        \textbf{Clock rate} & \multicolumn{4}{c}{10 GHz} & 5 GHz & 1.2 GHz &  1.28 GHz\\
        \textbf{Benchmark} & \multicolumn{2}{c}{AlexNet} & \multicolumn{2}{c}{ResNet-50} & AlexNet & ResNet-50 &  AlexNet \\
        \textbf{Batch size} & 1& 192 & 1& 58 & 1 & 1  & N/A \\
        \textbf{IPS} & 6,478 & 217, 201& 12,641 & 35,698 & 7,692 & 9,345 & 50,000 \\ 
        \textbf{IPS/W} & 872.17 & 7,476.78 & 1,021.17 & 1,587.99 & 344.17 & 100 & 900   \\ 
        \textbf{IPS/W/$\text{mm}^2$ }& 1.23 & 10.59 & 1.59 & 2.25 & 2.75 & 0.45 & 40.07  \\
        
\bottomrule
    \end{tabular}
\end{table*}

\subsubsection{Photonic Accelerators}

In Section~\ref{sec:bk-rel-wk}, using Table~\ref{table:photonic-acc} we discussed that previous works on photonic accelerators have not provided a full system evaluation. 
For completeness, in Table~\ref{table:sota-photonic-acc} we provide a quantitative comparison against the state-of-the-art photonic accelerators. 
The numbers reported in Table~\ref{table:sota-photonic-acc} are highly dependent on the various design choices, i.e. careful consideration of optical device choices, ADC/DAC choices, the on-chip memory sizes, non-linear units, communication links, etc., and the comprehensiveness of the evaluation.
Previous works use very small on-chip memory arrays (in KBs).
These small on-chip memory arrays have small area and power consumption, but require frequent DRAM transfers.
Not all previous works have considered this DRAM transfer overhead.  
When designing ADEPT, we considered the sizes of the weights and the activations of the neural networks. 
The high throughput goal of the system necessitates an adequately large SRAM array that enables the DNN inference to run without being bottlenecked by the off-chip data transfers. 
We can see that the full system of \ElectroPhotoAcc is not the most area efficient (due to large SRAM arrays), but it can provide 2.5 $\times$ better IPS/W than Albireo-C and 10.2 $\times$ better IPS/W than DNNARA for the same batch size of 1.
Although the batch size is not reported in HolyLight, ADEPT's and HolyLight's power efficiencies are comparable when ADEPT uses a batch size of 1. 
However, ADEPT's activation SRAM array is adequate to store even larger batch sizes which increases the utilization of the photo-core---providing a better overall system performance. 
When the maximum batch size is used for ADEPT, it can provide more than 8.3$\times$ better power efficiency compared to other three photonic accelerators.

\textbf{It should be noted that the goal of this paper is not to claim a more performant photonic core. 
In contrast, we aim to highlight the importance of a system-level analysis when evaluating photonic accelerators and encourage the community to adopt a pragmatic approach.}
The reasons that we achieve better results compared to other photonic accelerators despite their lack of system level analysis can be listed as: (1) we use low loss MEMS-based MZIs~\cite{noems} (0.04 dB) enabling a power efficient large (128 $\times$ 128) MZI mesh in ADEPT; (2) we use large SRAM arrays enabling large batch sizes and better utilization of the photo-core; and (3) the choice of data converters (ADCs/DACs) is different in different designs.

\section{Discussion}
\label{sec:discussion}

We sought to develop a balanced architecture that benefits from accelerating GEMM using photonics (1) without being bottlenecked by digital electronic operations or storage overhead, and (2) more than compensates for the overheads of electrical-optical and analog-digital conversions.
To this end, it is important to carefully formulate performance metrics to clearly see the system-level benefit of using photonics.
In particular, we use IPS as the throughput performance metric instead of TOPS.
The TOPS metric fails to consider processing unit utilization which is not likely to be unity.

In our proposed architecture, we perform electrical-optical and analog-digital conversions for input and output vectors each cycle. 
Although the overhead of performing conversions can improve with the process technology developments, it will remain a fundamental limitation for the speed and efficiency of the system. 
Hence, it may be worth performing more operations in the optical domain.
However, this increases losses in optical devices which lowers SNR and lowers bit precision (see Eq.~\eqref{eq:laser_power}). 
Similarly, the limited bandwidth of MZI leads to a 10 ns weight programming latency which limits system performance.
We minimize the impact of 10 ns MZI latency and power consumption of weight DACs by using a WS approach.

In the photo-core, the dynamic range of values are limited due to the output ADCs, which reduce the precision of MVM outputs (larger than 22 bit) back to 8-bit. In Section~\ref{sec:precision}, we discussed how to preserve 8-bit accuracy at the output vectors.
However, extra training efforts are still necessary to keep the accuracy in the desired range. 
We confirmed that 8-bit precision is sufficient to maintain the accuracy of the benchmarks we used (e.g. ResNet-50, BERT-large and RNN-T) within $1\%$ of the FP32 accuracy after performing several epochs of quantization-aware retraining~\cite{quantization-wu-2020, quantization-krishnamoorthi-2018}. 
While 8-bit precision is adequate for inference, training in the photo-core requires higher precision, which would lead to higher power (roughly scales with $2^B$ where $B$ is the number of bits). 
More intelligent training schemes may be needed to overcome this problem~\cite{swalp, training-high-acc}.

Our study shows that SRAM dominates the area of both ADEPT and electronic \SystolicArrays.
It is beneficial to have large local SRAMs to accommodate large batches of inputs.
However, SRAM size is limited in a chip-scale system. 
Therefore, scaling out to multiple chips is required to increase the SRAM cache size.
The problem of designing a scaled-out system with multiple chips with multiple photo-cores, mapping a DNN model onto the many accelerators, and orchestrating the communication between them is part of our future work.

Our results show that different types of DNNs exhibit different utilization behaviors due to the differing shapes of weight and input matrices as well as due to the differences in the networks ``GEMM-heaviness''.
Generally, the benefit we obtain from using the photonic core decreases for those networks with more non-GEMM operations.
It is possible to specialize the design of ADEPT by rearchitecting the photo-core to support different dataflows instead of GEMM; tailoring the digital electronic ASIC for a given set of operations; and choosing the SRAM sizes according to the network’s weight size and the optimal batch size. 

\section{Conclusion}
\label{sec:concl}

In this paper, we proposed and evaluated an end-to-end hybrid system for accelerating DNN inference containing a new electro-photonic accelerator called ADEPT. 
We showed that accelerating DNN inference with photonics requires tight interplay between the photonic compute units for GEMM operations and the electronic logic units for non-GEMM operations.
The result is a balanced electro-photonic system architecture that has a throughput that is similar to the throughput of a system utilizing the widely-accepted \SystolicArray architecture while consuming significantly less power.
With the introduced optimization methods for pipelining operations and data transfers, we showed that we can leverage the high throughput of the photonics GEMM accelerator without being bottlenecked by electronic units.
Overall, we are optimistic that photonic computing is nigh, and we are looking forward to the application of the technology in real-life. 
Given its advantage over purely electronic systems in terms of IPS/W or IPS/W$\cdot \text{mm}^2$, we are confident that the technology will find its rightful place within the Cambrian explosion of AI accelerators.

%%%%%%%%% -- BIB STYLE AND FILE -- %%%%%%%%
\bibliographystyle{IEEEtranS}
\bibliography{adept-micro-2022}
%%%%%%%%%%%%%%%%%%%%%%%%%%%%%%%%%%%%

\end{document}